\documentclass{article}

\usepackage[square,sort,comma,numbers]{natbib}

\usepackage[svgnames,dvipsnames]{xcolor}
\usepackage{amssymb}
\usepackage[T1]{fontenc}
\usepackage{mathtools}
\newcommand{\sfrac}[2]{\frac{#1}{#2}}
\usepackage{braket}
\usepackage{physics}
\usepackage{tikz}
\usepackage{xparse}
\usepackage{floatrow}
\usepackage{varwidth}
\usepackage[inline]{enumitem}
\usepackage{bm}
\usepackage[unicode]{hyperref}
\usepackage[margin=1.25in,headheight=13.6pt]{geometry}
\usepackage{tasks}
\usepackage[utf8]{inputenc}

\usepackage{amsthm}

\usepackage[noabbrev,capitalize,nameinlink]{cleveref}
\usepackage{changepage}
\usepackage[normalem]{ulem}
\usepackage{booktabs}
\usepackage{array}
\usepackage{graphicx}
\usepackage{soul}
\usepackage[shortcuts]{extdash}
\usepackage{appendix}
\usepackage{pbox}
\usepackage{sectsty}
\usepackage[nopatch=footnote]{microtype}
\usepackage{flafter}

\hypersetup{%
  colorlinks,
  bookmarksnumbered,
  pdfencoding=unicode,
  allcolors=DarkSlateGrey
}

\newcolumntype{L}[1]{>{\raggedright\let\newline\\\arraybackslash\hspace{0pt}}m{#1}}
\newcolumntype{C}[1]{>{\centering\let\newline\\\arraybackslash\hspace{0pt}}m{#1}}
\newcolumntype{M}[1]{>{\centering\let\newline\\\arraybackslash\hspace{0pt}$}m{#1}<{$}}
\newcolumntype{R}[1]{>{\raggedleft\let\newline\\\arraybackslash\hspace{0pt}}m{#1}}

\usepackage{setspace}
\usepackage{eqparbox, etoolbox}
\newlist{rdescription}{description}{1}
\AtBeginEnvironment{rdescription}{%
\setlist[rdescription]{leftmargin =\dimexpr\eqboxwidth{Des}+\labelsep}}%

\SetLabelAlign{parright}{\strut\smash{\parbox[t]\labelwidth{\raggedleft#1}}}

\usetikzlibrary{automata, positioning, decorations.pathmorphing, calc, quotes} 

\tikzset{
    every state/.append style={
        execute at begin node=$,
        execute at end node=$
    },
    initial text = 
}

\usepackage{thmtools}
\usepackage{thm-restate}
\declaretheorem{theorem}
\declaretheorem[sibling=theorem]{lemma,corollary}

\theoremstyle{definition}
\newtheorem{fact}[theorem]{Fact}
\newtheorem{definition}[theorem]{Definition}

\newtheorem{claim}{Claim}[theorem]
\newenvironment{subproof}[1][Subproof]{%
  \begin{proof}[#1]%
}{%
  \end{proof}%
}

\crefname{fact}{fact}{facts}
\Crefname{fact}{Fact}{Facts}

\crefalias{constraint}{equation}
\crefname{constraint}{constraint}{constraints}
\Crefname{constraint}{Constraint}{Constraints}
\creflabelformat{constraint}{#2\textup{(#1)}#3} 
\crefalias{sentence}{equation}
\crefname{sentence}{sentence}{sentences}
\Crefname{sentence}{Sentence}{Sentences}
\creflabelformat{sentence}{#2\textup{(#1)}#3}
\crefalias{expression}{equation}
\crefname{expression}{expression}{expressions}
\Crefname{expression}{Expression}{Expressions}
\creflabelformat{expression}{#2\textup{(#1)}#3}

\NewDocumentEnvironment{delineate}{m}{\textcolor{cyan!70!black!}{> > > > Begin: #1 > > > >}}{\textcolor{red!70!black!}{< < < < End: #1 < < < <}}

\mathchardef\mhyphen="2D

\DeclarePairedDelimiter{\ceil}{\lceil}{\rceil}

\DeclarePairedDelimiter\paren\lparen\rparen

\DeclarePairedDelimiter\sqparen{[}{]}
\DeclarePairedDelimiter\paramset{.}{|}

\newcommand{\given}{\ensuremath{\:\middle|\:}}

\DeclareMathOperator{\PR}{\mathbb{P}}
\newcommand{\prv}[1]{\ensuremath{\PR\sqparen*{#1}}}
\DeclareMathOperator{\pr}{Pr}

\newcommand{\justOH}{\ensuremath{\mathit{O}}}
\newcommand{\OH}[1]{\ensuremath{\justOH\paren*{#1}}}

\newcommand{\oh}[1]{\ensuremath{\mathit{o}\paren*{#1}}}

\makeatletter
\newcommand*{\IfItalicsTF}{%
  \ifx\f@shape\my@test@it
    \expandafter\@firstoftwo
  \else
    \expandafter\@secondoftwo
  \fi
}
\newcommand*{\my@test@it}{it}
\makeatother

\newcommand{\contextsensitivemathrm}[1]{\IfItalicsTF{\mathit{#1}}{\mathrm{#1}}}

\newcommand\machineformat[1]{\ensuremath{\contextsensitivemathrm{#1}}}
\newcommand\langclassformat[1]{\ensuremath{\mathsf{#1}}}
\newcommand\restrictionformat[1]{\ensuremath{\mathrm{#1}}}
\newcommand{\langformat}[1]{\ensuremath{\mathtt{#1}}}

\newcommand\co{\restrictionformat{con}}
\newcommand\lo{\restrictionformat{log}}

\newcommand\po{\restrictionformat{poly}}

\newcommand\rexlabel{\ensuremath{_\%}}
\newcommand\rex{\restrictionformat{\rexlabel}}
\newcommand\pexlabel{\ensuremath{^\%}}
\newcommand\pex{\restrictionformat{\pexlabel}}
\newcommand\pub{\restrictionformat{\mhyphen\allowbreak{}public}}
\newcommand\pri{\restrictionformat{\mhyphen\allowbreak{}private}}
\newcommand\spa{\restrictionformat{\mhyphen\allowbreak{}space}}
\newcommand\ran{\restrictionformat{\mhyphen\allowbreak{}coins}} 
\newcommand\tim{\restrictionformat{\mhyphen\allowbreak{}time}}

\newcommand{\IPhighlabel}{\langclassformat{IP}}
\newcommand{\IPlabel}{\langclassformat{IP_*}}

\newcommand{\IPhigh}[2][1]{\langclassformat{\IPhighlabel(\scalebox{#1}{$#2$})}}
\newcommand{\IP}[2][1]{\langclassformat{\IPlabel(\scalebox{#1}{$#2$})}}

\newcommand{\eL}[1]{\mathcal{L}\paren*{#1}}

\newcommand{\PP}{\langclassformat{P}}

\newcommand{\REG}{\langclassformat{REG}}

\newcommand{\NC}{\langclassformat{NC}}

\newcommand{\BQL}{\langclassformat{BQL}}
\newcommand{\BPL}{\langclassformat{BPL}}

\NewDocumentCommand{\DTIME}{ o m }{\langclassformat{DTIME\IfValueTF{#1}{\paren[#1]{#2}}{\paren*{#2}}}}

\newcommand{\BQTISPlabel}{\langclassformat{BQTISP_*}}
\newcommand{\BQTISP}[2][1]{\langclassformat{\BQTISPlabel(\scalebox{#1}{$#2$})}}

\newcommand{\BQTISPhighlabel}{\langclassformat{BQTISP}}
\newcommand{\BQTISPhigh}[2][1]{\langclassformat{\BQTISPhighlabel(\scalebox{#1}{$#2$})}}

\newcommand{\rat}{\ensuremath{\mathbb{Q}}}
\newcommand{\seqrat}{\langclassformat{S}^{=}_{\rat}}
\newcommand{\sneqrat}{\langclassformat{S}^{\neq}_{\rat}}

\newcommand{\dsfk}{DS-FK}

\NewDocumentCommand{\defineautomata}{ m m }{%
    \expandafter\NewDocumentCommand\csname#1fa\endcsname{}{\ensuremath{\machineformat{#1fa}}}%
    \expandafter\NewDocumentCommand\csname rt#1fa\endcsname{}{\ensuremath{\machineformat{rt\mhyphen#1fa}}}%
    \expandafter\NewDocumentCommand\csname o#1fa\endcsname{ o }{\ensuremath{\machineformat{1#1fa\IfValueT{##1}{\paren*{##1}}}}}%
    \expandafter\NewDocumentCommand\csname t#1fa\endcsname{ o }{\ensuremath{\machineformat{2#1fa\IfValueT{##1}{\paren*{##1}}}}}%
    \expandafter\NewDocumentCommand\csname#1tm\endcsname{}{\ensuremath{\machineformat{#1tm}}}%
    \expandafter\NewDocumentCommand\csname #2FA\endcsname{ d<> }{\ensuremath{\langclassformat{#2FA}\IfValueT{##1}{\paren*{##1}}}}%
    \expandafter\NewDocumentCommand\csname RT#2FA\endcsname{ d<> }{\ensuremath{\langclassformat{RT\mhyphen#2FA}\IfValueT{##1}{\paren*{##1}}}}%
    \expandafter\NewDocumentCommand\csname O#2FA\endcsname{ o d<> }{\ensuremath{\langclassformat{1#2FA}\IfValueTF{##1}{\paren*{##1\IfValueT{##2}{, ##2}}}{\IfValueT{##2}{\paren*{##2}}}}}%
    \expandafter\NewDocumentCommand\csname T#2FA\endcsname{ o d<> }{\ensuremath{\langclassformat{2#2FA}\IfValueTF{##1}{\paren*{##1\IfValueT{##2}{, ##2}}}{\IfValueT{##2}{\paren*{##2}}}}}%
}

\defineautomata{d}{D}
\defineautomata{n}{N}
\defineautomata{a}{A}
\defineautomata{p}{P}
\defineautomata{q}{Q}
\defineautomata{qc}{QC}

\newcommand{\ith}[2][th]{\ensuremath{#2}\text{#1}}

\newcommand\setneg[1]{\ensuremath{\overline{#1}}}

\newcommand{\PL}{\langformat{PAD}}

\newcommand{\SL}{\langformat{SEG}}

\newcommand{\langL}{\langformat{L}}
\newcommand{\langA}{\langformat{A}}
\newcommand{\langB}{\langformat{B}}
\newcommand{\pal}{\langformat{PAL}}
\newcommand{\twin}{\langformat{TWIN}}
\newcommand{\mult}{\langformat{MULT}}
\newcommand{\squarelang}{\langformat{SQUARE}}

\newcommand{\subsquare}{\langformat{SUBSQUARE}}
\newcommand{\power}{\langformat{POWER}}
\newcommand{\powereq}{\langformat{POWER\mhyphen{}EQ}}
\newcommand{\orderedeq}{\langformat{EQ_{ab}}}
\newcommand{\mixedeq}{\langformat{EQ}}

\newcommand\rev[1]{\ensuremath{#1^{R}}}

\newcommand{\acc}{\ensuremath{\mathrm{acc}}}
\newcommand{\rej}{\ensuremath{\mathrm{rej}}}

\newcommand{\sacc}{\ensuremath{s_{\acc}}}
\newcommand{\srej}{\ensuremath{s_{\rej}}}

\newcommand{\lend}{\ensuremath{\rhd}}
\newcommand{\rend}{\ensuremath{\lhd}}

\newcommand{\Accept}{\emph{Accept}}
\newcommand{\Reject}{\emph{Reject}}
\newcommand{\accept}{\emph{accept}}
\newcommand{\reject}{\emph{reject}}

\newcommand{\err}{\ensuremath{\varepsilon}}
\newcommand{\errprematuretimeout}{\ensuremath{\err_{\mathrm{premature}}}}

\newcommand{\verr}{\err}
\newcommand{\verracc}{\ensuremath{\verr^+}}
\newcommand{\verrrej}{\ensuremath{\verr^-}}
\newcommand{\verrloop}{\ensuremath{\verr^{\mathrm{loop}}}}

\setlist{itemsep=0pt}

\newlist{observations}{enumerate}{1}
\setlist[observations]{
    label=\arabic{*}.,
    ref=\arabic{*}
}
\crefname{observationsi}{observation}{observations}
\Crefname{observationsi}{Observation}{Observations}

\newlist{differences}{enumerate}{1}
\setlist[differences]{
    label=\arabic{*}.,
    ref=\arabic{*}
}
\crefname{differencesi}{difference}{differences}
\Crefname{differencesi}{Difference}{Differences}

\newlist{strategies}{enumerate}{1}
\setlist[strategies]{
    label=Strategy \arabic{*}:,
    ref=\arabic{*},
    labelwidth=\widthof{Strategy 1:},
    leftmargin=\parindent+\labelwidth+\labelsep
}
\crefname{strategiesi}{strategy}{strategies}
\Crefname{strategiesi}{Strategy}{Strategies}

\newlist{turingenum}{enumerate}{1}
\setlist[turingenum]{
    noitemsep,
    labelsep=.5em,
    leftmargin=1em+\parindent,
    labelwidth=1em,
    label=(\Roman{*}),
    ref=\Roman{*}
}

\crefname{turingenumi}{Stg.}{Stgs.}
\Crefname{turingenumi}{Stage}{Stages}

\SetLabelAlign{Center}{\hss#1\hss}

\newcommand{\narrowfont}[1]{\fontfamily{lmss}\fontseries{sbc}\selectfont #1}
\newcommand{\turinglabelformat}[1]{\narrowfont{\scriptsize#1}}
\newlength{\turinglabelgap}

\ExplSyntaxOn
\DeclareExpandableDocumentCommand{\IfNoValueOrEmptyTF}{mmm}
 {
  \IfNoValueTF{#1}{#2}
   {
    \tl_if_empty:nTF {#1} {#2} {#3}
   }
 }
\ExplSyntaxOff

\NewDocumentEnvironment{turing}{ O{} m m }
 {\IfNoValueOrEmptyTF{#1}{\setlength{\turinglabelgap}{0em}}{\setlength{\turinglabelgap}{0.5em}}\small\begin{enumerate}[labelsep=0pt,align=left,parsep=0pt,leftmargin=\widthof{\turinglabelformat{#1}}+\turinglabelgap, 
 listparindent=0pt] 
  \item[]\ignorespaces#3\\[0.5em]
  \begin{turingenum}[
    nosep,
    align=Center,
    labelwidth=\widthof{\turinglabelformat{#1}},
    labelsep=\turinglabelgap,
    leftmargin=0em 
  ]}
 {\unskip\end{turingenum}\end{enumerate}}

\usepackage{crossreftools}
\makeatletter
\newcommand{\optionaldesc}[3]{%
  \phantomsection
\protected@edef\@currentlabel{#1}\protected@edef\cref@currentlabel{%
    [#3][\arabic{#3}][\cref@result]%
    #1%
  }\label{#2}%
}
\makeatother

\NewDocumentCommand{\defineturingitem}{ m m }{%
    \expandafter\NewDocumentCommand\csname#1item\endcsname{ o o m }{\IfValueTF{##1}{\item[\turinglabelformat{##1}]\begin{adjustwidth}{#2}{0pt}\ignorespaces##3\end{adjustwidth}\optionaldesc{##1}{\IfValueTF{##2}{##2}{stg:##1}}{turingenumi}}{\item[]\begin{adjustwidth}{#2}{0pt}\ignorespaces##3\end{adjustwidth}}}%
}

\defineturingitem{t}{0em}
\defineturingitem{tt}{1em}
\defineturingitem{ttt}{2em}
\defineturingitem{tttt}{3em}
\defineturingitem{ttttt}{4em}

\newcommand{\msg}[1]{\raisebox{-1.5pt}{\fbox{\raisebox{1.5pt}{~#1~}}}}

\newcommand{\Yes}{\ensuremath{\mathrm{Yes}}}
\newcommand{\No}{\ensuremath{\mathrm{No}}}

\makeatletter
\def\squiggly{\bgroup \markoverwith{\lower3.9\p@\hbox{\sixly \scalebox{1.2}[0.65]{\char58}}}\ULon}
\makeatother

\def\mysout{\leavevmode\bgroup\def\ULthickness{1pt}\ULdepth=-.4ex\ULset}

\newcommand{\circlethickness}{7}
\newcommand{\circlesize}{0.5}
\newcommand{\circleopacity}{0.25}

\renewcommand{\circlethickness}{4}
\renewcommand{\circlesize}{0.2}
\renewcommand{\circleopacity}{0.2}

\newcommand{\stkout}[1]{\begingroup\ifmmode\text{\mysout{\ensuremath{#1}}}\else\mysout{#1}\fi\endgroup}
\newcommand{\utkanadd}[2][0]{\begingroup\color{blue!#1!cyan!70!black}\tikz[overlay]{\draw[color=blue!#1!cyan!50!black,line width=\circlethickness,opacity=\circleopacity] (0,0.1) circle (\circlesize);}#2\endgroup}

\newcommand{\utkanrem}[1]{\begingroup\color{red!80!black!70}\tikz[overlay]{\draw[color=red!50!black!70,line width=\circlethickness,opacity=\circleopacity] (0,0.1) circle (\circlesize);}\stkout{#1}\endgroup}
\newcommand{\utkanremlong}[2][\bigast\textbf{SUGGESTED FOR REMOVAL}\bigast]{\begingroup\color{red!80!black!70}#1 #2 #1\endgroup}

\newcommand{\utkanalt}[2]{\begingroup\color{green!45!black}\text{\texttt{\Large[}}#1\text{\texttt{\Large]}} \text{$\thickapprox$OR$\thickapprox$} \text{\texttt{\Large[}}#2\text{\texttt{\Large]}}\endgroup}

\newcommand{\utkanworry}[1]{\textcolor{red!45!black!90}{\ifmmode\smash[b]{\squiggly{#1}}\else\squiggly{#1}\fi}}

\renewcommand{\textvisiblespace}[1][.7em]{%
  \makebox[#1]{%
    \kern.07em
    \vrule height.5ex
    \hrulefill
    \vrule height.5ex
    \kern.07em
  }
}

\DeclareMathAlphabet{\mathsl}{\encodingdefault}{\familydefault}{m}{sl}
\SetMathAlphabet{\mathsl}{bold}{\encodingdefault}{\familydefault}{bx}{sl}

\usepackage{fancyhdr}
\author{A. C. Cem Say\\\href{mailto:say@bogazici.edu.tr}{\texttt{say@bogazici.edu.tr}}
\and
M. Utkan Gezer\\\href{mailto:utkan.gezer@bogazici.edu.tr}{\texttt{utkan.gezer@bogazici.edu.tr}}}
\title{Unconditional proofs of quantumness between
small-space machines}
\date{\small \itshape
  Department of Computer Engineering,
  Bo\u{g}azi\c{c}i University,
  Bebek 34342,
  \.{I}stanbul,
  T\"{u}rkiye
  }

\newcommand{\eg}{e.g.}

\newcommand{\ie}{i.e.}

\NewDocumentCommand{\padlang}{ e{_} m }{\ensuremath{\PL_{\IfValueT{#1}{#1}}\paren*{#2}}}
\newcommand{\padsymb}{\ensuremath{\star}}
\newcommand{\seglang}[1]{\ensuremath{\SL\paren*{#1}}}
\newcommand{\segsep}{\ensuremath{\#}}

\newcommand{\asd}{\ensuremath{\langclassformat{ASD}}}
\NewDocumentCommand{\padpair}{ s m }{%
    \IfBooleanTF{#1}%
    {\ensuremath{\paren*{\padlang{#2}, \padlang{\setneg{#2}}}}}%
    {\ensuremath{\paren{\padlang{#2}, \padlang{\setneg{#2}}}}}%
}

\NewDocumentCommand{\branch}{ m o }{\makebox{\textsc{[#1]}\IfValueT{#2}{ Branch #2:}}}
\NewDocumentCommand{\branchpr}{ m o }{\branch{\,Probability: \ensuremath{#1}\,}[#2]}
\NewDocumentCommand{\branchperc}{ m o }{\branch{#1\% prob.}[#2]}

\NewDocumentCommand{\vartextvisiblespace}{ O{.7em} O{.7ex} }{%
  \makebox[#1]{%
    \kern.07em
    \vrule height#2
    \hrulefill
    \vrule height#2
    \kern.07em
  }
}

\newcommand{\estring}{\ensuremath{\lambda}}
\newcommand{\blanksymb}{\ensuremath{\text{\vartextvisiblespace}}}
\NewDocumentCommand{\Qpub}{ e{_} }{\ensuremath{Q_{\textnormal{pub}\IfValueT{#1}{,#1}}}}
\NewDocumentCommand{\Qpri}{ e{_} }{\ensuremath{Q_{\textnormal{pri}\IfValueT{#1}{,#1}}}}
\NewDocumentCommand{\Qcom}{ e{_} }{\ensuremath{Q_{\textnormal{com}\IfValueT{#1}{,#1}}}}
\NewDocumentCommand{\bpub}{ e{_} }{\ensuremath{b_{\textnormal{pub}\IfValueT{#1}{,#1}}}}
\NewDocumentCommand{\bpri}{ e{_} }{\ensuremath{b_{\textnormal{pri}\IfValueT{#1}{,#1}}}}

\makeatletter

\def\@testdef #1#2#3{%
  \def\reserved@a{#3}\expandafter \ifx \csname #1@#2\endcsname
 \reserved@a  \else
\typeout{^^Jlabel #2 changed:^^J%
\meaning\reserved@a^^J%
\expandafter\meaning\csname #1@#2\endcsname^^J}%
\@tempswatrue \fi}

\makeatother

\begin{document}

\maketitle

\begin{abstract}
\noindent A proof of quantumness is a protocol through which a classical machine can test whether a purportedly quantum device, with comparable time and memory resources, is performing a computation that is impossible for classical computers. Existing approaches to provide proofs of quantumness depend on unproven assumptions about some task being impossible for machines of a particular model under certain resource restrictions. We study a setup where both devices have space bounds $\oh{\log \log n}$. Under such memory budgets, it has been unconditionally proven that probabilistic Turing machines are unable to solve certain computational problems. We formulate a new class of  problems, and show that these problems are  polynomial\-/time solvable for quantum machines, impossible for classical machines, and have the property that their solutions can be ``proved'' by a small\-/space quantum machine to a classical  machine with the same space bound. These problems form the basis of our newly defined  protocol, where the polynomial\-/time verifier's verdict about the tested machine's quantumness is not conditional on an unproven weakness assumption. 
  \vspace{1em}

\end{abstract}

\section{Introduction}\label{sec:intro}
It is hoped that sufficiently advanced quantum computers will be able to solve problems that are impossible for classical computers employing comparable amounts of resources. This brings up  interesting questions about how such a claim of ``quantum advantage'' can be verified by a classical machine: How can a device check the correctness of a claimed solution for a problem that is  too difficult for it to solve on its own? Furthermore, even if the ``weak'' device can verify that the problem has been solved correctly, is this sufficient to conclude that the ``strong'' device is indeed a quantum computer, rather than a classical one implementing a fast algorithm unknown to the verifying party?

All approaches to providing ``proofs of quantumness'' listed by Brakerski et al.~\cite{BKVV20}
depend on unproven assumptions about a particular task being difficult for computational machines of a particular model. As an example, in one such approach, a ``strong'' computer can factorize a given composite number,  the verifier can quickly check that the factors are correct, and then ``conclude'' that it is faced with a computer that has run an efficient quantum algorithm like~\cite{S97},
based on the belief that polynomial\-/time factorization is impossible for classical machines. 

More recently,   Zhan~\cite{Z23} and Girish et al.~\cite{GRZ24}
have focused on proofs between logarithmic\-/space bounded machines. They have shown that, for any language in the class \BQL{}, such a quantum machine can prove its work (on deciding about the membership of the common input string in the language) to a classical logspace verifier by transmitting  a ``proof string'' of polynomial length. Considered  as a framework for proofs of quantumness, this approach is also based on the (as yet unproven) assumption that \BQL{} is a proper superset of \BPL{}, the corresponding class for classical logspace machines.

In this paper, we present the first framework for proofs of quantumness where the conclusion of the verifier is not conditional on an unproven presupposition about a weakness of a family of machines, by studying the case where both interacting computers have drastically small ($\oh{\log \log n}$) memory budgets. It is known~\cite{DS92,FK94} that certain separation problems are impossible for
probabilistic Turing machines (\ptm's) using $\oh{\log \log n}$ space, even if they are allowed to employ uncomputable numbers as their transition probabilities. On the other hand,  
two\-/way finite automata with quantum and classical states (\tqcfa{}'s), which are, essentially, constant\-/space Turing machines which have been augmented with a fixed\-/size quantum register, are known~\cite{AW02,SY17,R20,AY21}
to be able to perform some language recognition tasks that are beyond the capabilities of classical small-space machines.  We  combine these facts with techniques stemming from the study of interactive proof systems with finite\-/state verifiers~\cite{DS92,SY14,GS22,GDES23,GS24} and quantum finite automata~\cite{YS1009,YS1001} to construct setups where a polynomial\-/time interaction between the two space\-/bounded automata ends with the verifier accepting with very high probability  only if it is faced with a genuinely quantum machine obeying the communication protocol.
 
The paper is structured as follows: \Cref{sec:defs} presents the building blocks to be used to construct our framework. The small\-/space Turing machine models that  represent the classical and quantum computers which  take part in the tests of quantumness are defined.   We allow our classical machines to use the largest realistic set of transition probabilities, namely, computable reals between $0$ and $1$, and naturally impose the same computability restriction on the transition amplitudes of the quantum machines.\footnote{The \tqcfa{} algorithms we will use are known~\cite{AW02} to rely on the highly precise manipulation of a small number of qubits. This requirement is mitigated by our imposition of polynomial limits on their runtimes, noting that a polynomial\-/time quantum computation can be carried out within constant accuracy if the transition amplitudes are specified to logarithmically many bits of precision~\cite{BV93}. Our definitions allow the descriptions of classical machines that take part in our protocols to specify transition probabilities with the same precision as the quantum ones.} 
A list of tasks that are known or presumed to be difficult for small\-/space \ptm{}'s, but are solvable by \tqcfa{} algorithms, is given in \Cref{subs:pfaqfa}.  \Cref{subs:ips} describes how we pair a classical verifier with a (quantum or classical) ``prover'' in our protocols, and presents a method by which a constant\-/space prover can convince a similarly bounded verifier about the membership of their common input string in any language recognized by two\-/way deterministic two\-/head finite automata possessing a property that we define. To our knowledge, this is the first step in the characterization of the power of interactive proof systems with such severely constrained provers.

We give a formal definition of what we mean by ``checking a proof of quantumness'' in \Cref{sec:PoQdef}. Intuitively, one describes a verifier $V$ which accepts with high probability if it is faced with a genuinely quantum prover $P^Q$ that can convince $V$ that it is able to solve any given instance of a computational problem $\mathbf{P}$, and every classical prover $P^C$ that attempts to convince  $V$ is guaranteed to be  less successful than $P^Q$ on an infinite set of common input strings $W_{P^C}$. Constructing such a verifier therefore requires identifying a problem $\mathbf{P}$ that has the following properties:
\begin{enumerate}
    \item $\mathbf{P}$ should be solvable for a polynomial\-/time \qtm{}.
    \item $\mathbf{P}$ should be provably impossible to solve for a \ptm{} within the same time and space bounds.
    \item The solution of each instance of $\mathbf{P}$ should be ``provable'' by a \qtm{} to a \ptm{}, both operating within the same time and space bounds, through a polynomial\-/time interaction where only classical messages are exchanged.
\end{enumerate}
As  noted above, we are able to identify problems that satisfy all these properties where the common space bound is very small. In \Cref{sec:padded}, we present a specific family  of such problems and describe a template for proof\-/of\-/quantumness protocols for each member of that family.

In \Cref{sec:squ}, we give another protocol based on a problem that seems to be beyond the capability of the provers described in \Cref{sec:padded}. That verifier, which employs a version of Freivalds' celebrated (exponential\-/time) probabilistic finite\-/state algorithm~\cite{F81} for recognizing a nonregular language, can be considered to be checking a proof of quantumness on the condition that the underlying problem is classically difficult.  \Cref{sec:conc} lists several remaining open questions. The \nameref{sec:generalizedDS} presents a generalization of a theorem by Dwork and Stockmeyer~\cite{DS90} that may provide a starting point for future proof\-/of\-/quantumness protocols.


\section{Building blocks}\label{sec:defs}

We start by establishing some key strengths and weaknesses of the  machine models  that underlie  the ``proof of quantumness'' setup to be studied. \Cref{subs:pfaqfa} presents the ``stand\-/alone'' versions of the classical (probabilistic) and quantum  machines that will be relevant in our discussion, and lists the quantum advantage results under the considered common space bounds. 
In  \Cref{subs:ips}, we describe how those stand\-/alone machines can be put together to form interactive proof systems, and present a technique that will be used by the provers to be presented in \Cref{sec:padded}.

In the following, 
we assume that the reader is familiar with the fundamental notions of quantum computation~\cite{NC00}. Let $\mathcal{C}_{\sqparen*{0,1}}$  denote the set of computable numbers in $\sqparen*{0,1}$.

\subsection{Basic models and quantum advantage}\label{subs:pfaqfa}

The two basic computational models considered in this paper are  probabilistic and quantum Turing machines. In both cases, the machine has a read\-/only input tape. For any input string $w$, this tape will contain the string $\lend w \rend$, where $\lend$ and $\rend$ are special endmarker symbols. Details about the work tapes differ among models, as will be seen.

\begin{definition}\label{def:oncekisubsectiondakidef}
A \emph{probabilistic Turing machine (\ptm{})}\footnote{This definition has been obtained by restricting the transition probabilities of the model in~\cite{DS90} to computable numbers.}  is a 7\=/tuple  $\paren*{S,\Sigma,K,\delta,s_0,\sacc,\srej}$, where
\begin{itemize}
    \item $S$ is the finite set of states,
    \item $\Sigma$ is the finite input alphabet, not containing  $\lend$ and $\rend$, 
    \item $K$ is the finite work alphabet, including the blank symbol \blanksymb,
    \item 
    $\delta: \paren*{S \setminus \Set{\sacc,\srej}} \times \paren*{\Sigma \cup \Set{\lend, \rend}} \times K \times S \times K \times \Set{-1,0,1} \times \Set{-1,0,1} \to \mathcal{C}_{\sqparen*{0,1}}$, such that, for each fixed $s$, $\sigma$, and $\kappa$, 
    $\sum_{s', \kappa', d_i, d_w} \delta(s,\sigma,\kappa,s',\kappa',d_i,d_w)=1$, is the transition function,
    \item $s_0 \in S$ is the start state, and
    \item $\sacc,\srej \in S$, such that $\sacc \neq \srej$, are the accept and reject states, respectively.
\end{itemize}
\end{definition}

A \ptm{} has a single read\-/write work tape, which is initially blank, with the work head positioned on the leftmost cell. The computation of a \ptm{} $M$ on input $w \in \Sigma^*$ begins with the input head positioned on the first symbol of   the string $\lend w \rend$, and the machine in state $s_0$. In every computational step, if $M$ is in some state $s  \notin \Set{\sacc,\srej}$ with the input head scanning some symbol $\sigma$ at the \ith{i} tape position and the work head scanning some symbol $\kappa$ at the \ith{j} tape position, it switches to state $s'$, sets the scanned work symbol to $\kappa'$, and locates the input and work heads at the \ith{(i+d_i)} and \ith{(j+d_w)} positions respectively with probability $\delta(s,\sigma,\kappa,s',\kappa',d_i,d_w)$.
We assume that $\delta$ is defined such that the input head never moves beyond the tape area delimited by the endmarkers. If $M$ is in state $\sacc$ ($\srej$), it halts and accepts (rejects).

A \ptm{} is said to \emph{run within space} $s(n)$ if, on any input of length $n$, at most $s(n)$ work tape cells are scanned throughout the computation. We are going to focus on machines that run within space bounds that are $\oh{\log \log n}$. A computation performed by a machine that runs within $\OH{1}$ space can also be performed by a machine which does not use its work tape at all, and encodes all the workspace in its state set. The definition of this constant\-/space model, known as the \emph{two\-/way probabilistic finite automaton (\tpfa{})}, is easily obtained from \Cref{def:oncekisubsectiondakidef} by removing the components related to the work tape. A \emph{two\-/way deterministic finite automaton (\tdfa{})} is simply a \tpfa{} whose transition function is restricted to the range $\Set{0,1}$.

Our protocols in \Cref{sec:padded,sec:squ} will make use of the well\-/known capability of \tpfa{}'s to implement ``alarm clocks'' that can be set to any desired polynomial time bound. A detailed proof of the following fact can be found, for example, in~\cite{GS24}:

\begin{fact}\label{fact:factclock}
    For any pair of positive integers $c,t > 0$ 
    and desired ``error'' bound $\errprematuretimeout > 0$, there exists a \tpfa{} with expected runtime in \OH{n^{t+1}}, such that the probability that this machine halts in fewer than $c n^t$ steps is \errprematuretimeout.
\end{fact}

For our \emph{quantum Turing machine (\qtm{})} model, we adopt the definition of Watrous~\cite{W03}.
This model is best visualized as  a deterministic Turing machine (with a classical work tape) which has been augmented by adding a finite quantum register and a quantum work tape, each cell of which holds a qubit. The randomness in the behavior of a \qtm{} is a result of the operations performed on the quantum part of the machine at intermediate steps. A \qtm{} is said to run within space $s(n)$ if, on any input of length $n$, at most $s(n)$ cells are scanned on both the classical and quantum work tapes during the computation. 

It turns out that all the \qtm{}'s that we need in our results in this paper are constant\-/space machines. As described above, machines obeying this restriction can be modeled in a simpler way by  removing the work tapes from the definition. Doing this to the \qtm{}'s of~\cite{W03} yields the well\-/studied  quantum finite automaton model  due to Ambainis and Watrous~\cite{AW02}, which we describe in detail below.\footnote{Remscrim~\cite{R21} provides a detailed explanation of how quantum finite automata are generalized to \qtm{}'s with larger space bounds.}   The following definition, which can be viewed as a \tdfa{} augmented with a quantum register of constant size, is somewhat more involved than \Cref{def:oncekisubsectiondakidef}, because of the way it breaks down each computational step into two stages for evolving the quantum state and the remaining classical part of the machine separately:

\begin{definition}\label{def:ONCEKISEC2QCFADEF}
A \emph{two\-/way  finite automaton with quantum and classical states (\tqcfa)}  is a 9\=/tuple  $\paren*{Q,S,\Sigma,\delta_q,\delta_c,q_0,s_0,\sacc,\srej}$, where
\begin{itemize}
    \item $Q$ is the finite set of basis states of the quantum part of the machine,
    \item $S$ is the finite set of classical states,
    \item $\Sigma$ is the finite input alphabet, not containing  $\lend$ and $\rend$, 
    \item $\delta_q$ is the quantum transition function, which, for any $s \in (S \setminus \Set{\sacc,\srej})$ and $\sigma\in (\Sigma \cup \Set{\lend,\rend})$, maps the pair $(s,\sigma)$ to an action that will be performed on the quantum part of the machine, as will be described below,
    \item $\delta_c$ is the classical transition function, which determines the input head movement direction and resulting classical state associated with the current computational step, as will be described below,
    \item $q_0 \in Q$ is the initial quantum  state,
    \item $s_0 \in S$ is the classical start state, and
    \item $\sacc,\srej \in S$, such that $\sacc \neq \srej$, are the accept and reject states, respectively.
\end{itemize}
\end{definition}

Just like in the  classical case, the input string to a \tqcfa{} is delimited by endmarkers (beyond which the machine does not attempt to move its head) on the read\-/only tape. At the start of the computation, the input head is located on the left endmarker, the state of the quantum register is described by the vector $\ket{q_0}$, and the classical state is $s_0$.  Every computational step begins with the action  $\delta_q(s,\sigma)$ determined by the current classical state $s$ and the currently scanned tape symbol $\sigma$ being performed on the quantum register. Each such action is either a unitary transformation or a projective measurement of the quantum state, resulting in some outcome $\tau$ with an associated probability. The second and final stage of the computational step is a classical transition that determines the pair $(s',d)$ of the resulting classical state and head movement direction of this step: If  $\delta_q(s,\sigma)$ was a unitary transformation, this pair is simply  $\delta_c(s,\sigma)$, depending again only on $s$ and $\sigma$. If, on the other hand, $\delta_q(s,\sigma)$ was a measurement with outcome $\tau$,   the pair $(s',d)$ is  $\delta_c(s,\sigma,\tau)$, which also depends on that outcome. Computation halts with acceptance (rejection) when the \tqcfa{} reaches the classical state $\sacc$ ($\srej$).  

Since we want our classical and quantum machine models to be ``comparable'' in all respects except the fact that one is classical and the other is quantum, we restrict all entries of all unitary matrices describing the transitions of \tqcfa{}'s to be complex numbers whose real and imaginary parts are computable.\footnote{The unrealistic case where uncomputable numbers are allowed in the descriptions of these machines has been studied in~\cite{DS90,SY17}. Remscrim~\cite{R20} has investigated the effects of imposing more restrictions on the  set of allowed matrix entries.}

Let $\langA, \langB \subseteq \Sigma^*$ with $\langA \cap \langB =\emptyset$. We say that a (classical or quantum) machine $M$ \emph{separates $\langA$ and $\langB$ (with error bound $\err$)} if there exists a number $\err < \frac{1}{2}$ such that
\begin{itemize}
    \item for all $w \in \langA$, $M$ accepts input $w$ with probability at least $1-\err$, and
    \item for all $w \in \langB$, $M$ rejects input $w$ with probability at least $1-\err$.
\end{itemize}
Equivalently, one says that \emph{$M$ solves the promise problem $(\langA,\langB)$.} We say that $M$ \emph{recognizes $\langA$} if $M$ separates $\langA$ and the complement of $\langA$.

For any given \tpfa{} $M^{C}$ recognizing some language $\langL$, it is trivial to  construct a \tqcfa{} that recognizes $\langL$ by mimicking $M^{C}$ step by step. \tqcfa{}'s are, in fact, strictly more powerful computational machines than \tpfa{}'s. We review the relevant facts below.

Although some nonregular languages like $\orderedeq{}=\Set{a^n b^n | n\geq 0}$ are known~\cite{F81}  to have exponential\-/time \tpfa{}'s that recognize them, 
no $\oh{\log \log n}$\-/space \ptm{} can recognize a nonregular language in polynomial expected time, as proven by Dwork and Stockmeyer~\cite{DS90}. On the other hand,  Ambainis and Watrous~\cite{AW02} describe a polynomial\-/time \tqcfa{} that recognizes $\orderedeq{}$. Naturally, polynomial\-/time \tqcfa{}'s also exist for infinitely many other languages whose recognition is reducible to   determining whether two different symbols appear equally many times 
in the input string, \eg, $\mixedeq = \Set{w | w\in \Set{a,b}^* \text{ and } w \text{ contains equally many $a$'s and $b$'s}}$~\cite{AY21} and
\begin{equation*}
    \powereq = \Set{ aba^{7}ba^{7 \cdot 8}ba^{7 \cdot 8^{2}} \dotsm ba^{7 \cdot 8^{n}} | n \geq 0 },
\end{equation*}
studied in~\cite{SY17}. More recently, Remscrim~\cite{R20} has shown that any language corresponding to the word problem of a finitely generated virtually abelian group can be recognized in cubic time by some bounded\-/error \tqcfa{}. $\mixedeq{}$ is one of the ``simplest'' members of this set of languages, whose more ``complicated'' members have alphabets of the form $\Set{a_1,b_1,a_2,b_2,\dotsc,a_k,b_k}$ and require the strings to contain equal numbers of $a_i$'s and $b_i$'s for every $i$. 
Let $\BQTISP{t(n),s(n)}$ denote the class of languages recognizable  by \qtm{}'s that run in at most $t(n)$ expected time and within at
 most $s(n)$ space on all inputs of length at most $n$, so that such a \qtm{} recognizing the corresponding language exists for any desired positive error bound. All languages mentioned in this paragraph are known to be in $\BQTISP{n^k,c}$ for some constants $k$ and $c$ depending on the specific language.

In their seminal paper, Ambainis and Watrous~\cite{AW02} also gave an exponential\-/time \tqcfa{} algorithm for \pal{}, the language of palindromes on a binary alphabet. This provided another early example of quantum superiority over classical machines, since it is known~\cite{DS92,FK94} that no sublogarithmic\-/space \ptm{} can recognize $\pal{}$ in any amount of time.

A \emph{rational \opfa{}}~\cite{T69} is a \tpfa{} whose transition function  is restricted to contain only rational numbers in its range, and not to assign positive probabilities to transitions that do not move the head to the right. 
A language $\langL$ is said to belong to the class $\seqrat$~\cite{T69,M93}
if there exists a rational \opfa{} that
accepts all and only the members of $\langL$ with probability $\frac{1}{2}$. $\sneqrat$ is the class of languages whose complements are in $\seqrat$.

Using a technique developed in~\cite{YS1009}, Yakary{\i}lmaz and Say  have proven~\cite{YS1001} that all languages in  $\seqrat \cup \sneqrat$ can be recognized  by \tqcfa{}'s~\cite{AY21}. Combining this result with Remscrim's work on word problems~\cite{R20}, \Cref{tab:languagesrecognizedinexp} lists some other interesting languages known to have \tqcfa{}'s that recognize them  in $2^{\OH{n}}$ expected time for inputs of length  $n$.\footnote{$w^R$ denotes the reverse of string $w$.} For each of those languages, and for any desired small positive value of $\varepsilon$, there exists a $2^{kn}$\=/time \tqcfa{} (with a correspondingly large $k$) recognizing the language with  error bound $\varepsilon$.

\begin{table}[H]
    \caption{Some languages recognized in $2^{\OH{n}}$ time by \tqcfa{}'s.}
    \label{tab:languagesrecognizedinexp}
    \begin{tabular}{C{5.2em} | L{35.7em}}
        \textbf{Language} & \textbf{Description}\\\toprule
        \pal & $\Set{w | w = w^R}$\\\midrule
        \twin & $\Set{w\#w | w \in \Set{a,b}^*}$\\\midrule
        \mult & $\Set{x\#y\#z | x, y, z \text{ are natural numbers in binary notation and } x \cdot y = z}$\\\midrule
        \squarelang & $\Set{a^i b^{i^2} | i > 0}$\\\midrule
        \power & $\Set{a^i b^{2^i}| i > 0}$\\\midrule
        any ``polynomial language'' & A \emph{polynomial language}~\cite{Tur82} is defined as \begin{equation*}\Set{ a_1^{n_1} \dotsm a_k^{n_k} b_1^{p_1(n_1,\dotsc,n_k)} \dotsm b_r^{p_r(n_1,\dotsc,n_k)} | p_i(n_1, \dotsc, n_k) \geq 0 },\end{equation*} where $a_1, \dotsc, a_k, b_1, \dotsc, b_r$ are distinct symbols, and each $p_i$ is a polynomial with integer coefficients.\\\midrule
        $\langformat{W_G}$ & the word problem~\cite{YS1009} for $G$, where $G$ is any finitely generated virtually free group
    \end{tabular}
\end{table}



\subsection{Interactive proof systems with small-space provers}\label{subs:ips}

An \emph{interactive proof system} consists of two entities called the \emph{verifier} and the \emph{prover} that share a common finite input alphabet $\Sigma$ and a finite communication alphabet $\Gamma$. 
The verifier definition in our setup will essentially be a version of \Cref{def:oncekisubsectiondakidef} that has been augmented to enable the machine to communicate with the prover.


\newcommand{\silent}{\ensuremath{\mathrm{silent}}}
\newcommand{\com}{\ensuremath{\mathrm{com}}}

\begin{definition}\label{def:BUSECVERIFIERDEF}
A \emph{verifier} 
is a 9\=/tuple  
$\paren*
{S_{\silent},S_{\com},\Sigma,K,\Gamma,\delta,s_0,\sacc,\srej}$, where
\begin{itemize}
    \item $S_{\silent}$ is the finite set of noncommunicating states,
    \item $S_{\com}$ is the finite set of communicating states, such that  $S_{\silent} \cap S_{\com} = \emptyset$,
    \item $\Sigma$ is the finite input alphabet, not containing  $\lend$ and $\rend$, 
        \item $K$ is the finite work  alphabet, including the blank symbol \blanksymb,
        \item $\Gamma$ is the finite communication alphabet, 
    \item $\delta$  is the transition function, described below,
    \item $s_0 \in S_{\silent}$ is the start state, and
    \item $\sacc,\srej \in S_{\silent}$, such that $\sacc \neq \srej$, are the accept and reject states, respectively.
\end{itemize}
\end{definition}


The main difference between this verifier model and the ``stand\-/alone'' \ptm{} model of \Cref{def:oncekisubsectiondakidef} is the communication cell, through which the messages to and from the prover will be conveyed. When a verifier $V$ starts execution, the communication cell contains the blank symbol \blanksymb.
The transition function $\delta$ is defined in two parts. In any computational step starting with $V$ at some state $s \in S_{\silent}$ with its input head scanning some symbol $\sigma$ at the \ith{i} tape position, its work head scanning some symbol $\kappa$ at the \ith{j} tape position, and the communication cell containing the symbol $\gamma$, $\delta(s,\sigma,\kappa,\gamma,s',\kappa',d_i,d_w)$ is the probability with which $V$ will switch to state $s'$, write $\kappa'$ in the scanned work tape cell,  and then locate the input and work heads at the \ith{(i+d_i)} and \ith{(j+d_w)} tape positions, respectively.

Each communication state $s$ is associated with a symbol $\gamma_s \in \Gamma$. 
Whenever the machine enters a state $s \in S_{\com}$, the symbol $\gamma_s$ is written in the communication cell. As will be described below, the prover is supposed to respond to this communication with some symbol $\gamma'$ in its next computational step.
If this response is not received by the time $V$ starts its next step, it rejects immediately.
Otherwise, $V$ transitions from state $s$ to state $s'$, replaces the scanned work tape symbol with $\kappa'$, and moves its heads in the directions associated with $d_i$ and $d_w$, respectively, with probability $\delta(s,\sigma,\kappa,\gamma',s',\kappa',d_i,d_w)$, having received the communication symbol $\gamma'$ in response to its message while  scanning the input symbol $\sigma$ and the work tape symbol $\kappa$.

Note that both the runtime and the acceptance probability of a verifier  reading a particular input string and communicating with a particular prover may depend on the messages of the prover provided as responses during this communication.

In the study of interactive proof systems where computationally limited verifiers are faced with provers of unbounded power, one  has no need to commit to a specific machine definition for the role of the prover. In the context of this paper, the prover is known to be a $\oh{\log \log n}$\-/space machine, and it is the verifier's task to determine whether the prover is indeed quantum (\ie, a \qtm{} augmented with a communication cell) as it claims to be, or just a  classical probabilistic Turing machine as the verifier itself. In the following  definitions of these two types of provers, we obey the convention~\cite{DS92} that the prover sends a symbol through the communication cell only at the steps in which it detects that a new communication symbol has just been written by the verifier. As mentioned above, the \qtm{} algorithms we will describe require only $\OH{1}$ space, so our definition of a quantum prover is just a properly augmented version of \Cref{def:ONCEKISEC2QCFADEF}:

\begin{definition}\label{def:qprover}
A \emph{(constant\-/space) quantum prover}  is a 9\=/tuple  $\paren*{Q,S,\Sigma,\Gamma,\delta_q,\delta_{\silent},\delta_{\com},q_0,s_0}$, where
\begin{itemize}
    \item $Q$ is the finite set of basis states of the quantum part of the machine,
    \item $S$ is the finite set of classical states,
    \item $\Sigma$ is the finite input alphabet, not containing  $\lend$ and $\rend$, 
    \item $\Gamma$ is the finite communication alphabet, 
    \item $\delta_q$ is the quantum transition function, which, for any $s \in (S \setminus \Set{\sacc,\srej})$ and $\sigma\in (\Sigma \cup \Set{\lend,\rend})$, maps the pair $(s,\sigma)$ to an action (a unitary transition or a measurement) that will be performed on the quantum part of the machine, exactly as  in \Cref{def:ONCEKISEC2QCFADEF},
    \item $\delta_{\silent}$ is the ``silent'' classical transition function, which determines the input head movement direction and resulting classical state associated with the current computational step when no new incoming symbol has been detected in the communication cell, as the function $\delta_c$ in \Cref{def:ONCEKISEC2QCFADEF},
    \item $\delta_{\com}$ is the classical transition function to be employed when a new message from the verifier has been detected in the communication cell, as described below,
    \item $q_0 \in Q$ is the initial quantum  state, and
    \item $s_0 \in S$ is the classical start state.
\end{itemize}
\end{definition}

Any such quantum prover $P^Q$ has access to a communication cell that it shares with the verifier. Just like the stand\-/alone \tqcfa{}'s we saw in \Cref{subs:pfaqfa}, $P^Q$ performs each computational step in two stages. In addition to the quantum and classical transition functions $\delta_q$ and $\delta_{\silent}$, which serve the same purpose as their counterparts in \Cref{def:ONCEKISEC2QCFADEF}, $P^Q$ also has a second classical transition function $\delta_{\com}$, which is employed instead of $\delta_{\silent}$ only at the steps where a new communication symbol written by the verifier is detected in the communication cell. Each such step starting from classical state $s$ and with symbol $\sigma$ being scanned in the input tape shared by the verifier first performs the quantum transition indicated by $\delta_q(s,\sigma)$ as described in \Cref{subs:pfaqfa}, but continues with a classical transition (that also depends on the symbol $\gamma$ in the communication cell) from $\delta_{\com}$, rather than $\delta_{\silent}$. In more detail, if $\delta_q(s,\sigma)$ was a unitary transition, $\delta_{\com}(s,\sigma,\gamma)$ is a triple $(s',d,\gamma')$, specifying the resulting classical state $s'$ and input head direction $d$ of $P^Q$,  as well as the  symbol $\gamma'$ that $P^Q$ will write in the communication cell, overwriting its previous content. If $\delta_q(s,\sigma)$ dictated a measurement with outcome $\tau$, $\delta_{\com}(s,\sigma,\tau,\gamma)$ is similarly  a triple indicating the new state, head direction, and response of $P^Q$. 

We define a classical prover analogously. Since we will  aim to prove that the verifiers to be presented in \Cref{sec:padded,sec:squ} reject the ``quantumness'' claims of \emph{any} classical  prover whose space bound  is (a possibly increasing function of $n$) within $\oh{\log \log n}$, we base this model on the  general \ptm{} with the work tape (\Cref{def:oncekisubsectiondakidef}).

\begin{definition}\label{def:cprover}
A \emph{classical prover}  is a 7\=/tuple  $\paren*{S,\Sigma,\Gamma,K,\delta_{\silent},\delta_{\com},s_0}$, where
\begin{itemize}
    \item $S$ is the finite set of states,
    \item $\Sigma$ is the finite input alphabet, not containing  $\lend$ and $\rend$,
    \item $K$ is the finite work alphabet, containing the blank  symbol \blanksymb,
    \item $\Gamma$ is the finite communication alphabet, 
    \item $\delta_{\silent}: \paren*{S \setminus \Set{\sacc,\srej}} \times \paren*{\Sigma \cup \Set{\lend, \rend}} \times K \times S \times K \times \Set{-1,0,1} \times \Set{-1,0,1} \to \mathcal{C}_{\sqparen*{0,1}}$, such that, for each fixed $s$, $\sigma$, and $\kappa$, 
    $\sum_{s', \kappa', d_i, d_w} \delta(s,\sigma,\kappa,s',\kappa',d_i,d_w)=1$, is the ``silent'' transition function, as in \Cref{def:oncekisubsectiondakidef},
    \item $\delta_{\com}: \paren*{S \setminus \Set{\sacc,\srej}} \times \paren*{\Sigma \cup \Set{\lend, \rend}} \times K \times \Gamma \times S \times K \times \Gamma \times \Set{-1,0,1} \times \Set{-1,0,1} \to \mathcal{C}_{\sqparen*{0,1}}$, such that, for each fixed $s$, $\sigma$, $\kappa$, and $\gamma$, 
    $\sum_{s', \kappa', \gamma', d_i, d_w} \delta(s,\sigma,\kappa,\gamma,s',\kappa',\gamma',d_i,d_w)=1$, is the  transition function to be employed at steps where a fresh message from the verifier is detected in the communication cell, as described below, and
    \item $s_0 \in S$ is the start state.
\end{itemize}
\end{definition}

The difference between \Cref{def:BUSECVERIFIERDEF,def:cprover}, which both describe \tpfa{}'s augmented with a communication cell, is caused by the need to follow the convention that a prover only ``talks'' in response to a communication from the corresponding verifier. As in \Cref{def:qprover}, we specify two different transition functions, $\delta_{\silent}$ and $\delta_{\com}$: $\delta_{\com}(s,\sigma,\kappa,\gamma,s',\kappa',\gamma',d_i,d_w)$ is the probability with which the classical prover will switch to state $s'$, write $\kappa'$ on the work tape, write $\gamma'$ in the communication cell, and move the input and work tape heads in directions $d_i$ and $d_w$, respectively,  at a step where it has detected a fresh communication symbol $\gamma$ while it is in state $s$ and its input and work heads are scanning the symbols $\sigma$ and $\kappa$, respectively. $\delta_{\silent}$, which is identical in function to its counterpart in \Cref{def:oncekisubsectiondakidef}, is employed at steps where no ``news'' has been received from the verifier.

For any verifier $V$ and prover $P$ running on a common input string $w$, we let $\pr_{\acc}(V,P,w)$ (resp.\ $\pr_{\rej}(V,P,w)$) denote the probability that $V$ accepts (resp.\ rejects) and halts as a result of the interaction. We will also be considering the possibility that $V$ is ``tricked'' into running forever without reaching a halting state.

In a standard interactive proof system (IPS) as in~\cite{Sha92}, the prover's aim is to  convince the verifier to accept with high probability that their common input string is a member of some language $\langL$, and we say that $\langL$ \emph{has an IPS with error bound $\err$} if there exists a  verifier that succeeds in making use of the communication with the prover to accept every member of $\langL$, \emph{and} avoiding being tricked into looping on or accepting any non\-/member of $\langL$ with  probability at least $1-\err$, for some $\err<\frac{1}{2}$. In the next section, we will show  how to modify this setup so that the end result of the dialogue between the verifier and the prover can be interpreted as a verdict about whether the prover is a quantum machine or not. In the remainder of this section, we  present a technique that can be used by a small\-/space prover faced with a similarly bounded verifier in a standard IPS. This technique will be employed by the quantum provers to be described in Section \ref{sec:padded}. 

A \emph{two\-/head two\-/way deterministic finite automaton (\tdfa[2])}~\cite{HKM11} can be visualized as a \tdfa{} with not one, but two heads. Both heads are positioned on the left endmarker at the beginning of the computation. The transition function governs the two heads separately. At any step of the computation, the next move (\ie, the next state and the movement directions of the two heads) is determined by the current internal state and the symbol pair scanned by the two heads. The class of languages that can be recognized by \tdfa[2]'s, designated by \TDFA[2], is known to contain many interesting nonregular languages. We will concentrate on a subclass associated with a more restricted machine definition.

In the following, we will construct a verifier algorithm for simulating a given \tdfa[2] $M$ on a given input. At each step of the simulation, the verifier, which has only one input head, will receive the  symbol which is purportedly scanned by one of the heads of $M$ from the prover. This will give the prover a chance to ``lie'' about the reading of that head in order to trick the verifier to accept the input string, even if it is not a member of the language recognized by $M$. We now define a property of heads that will be important in this regard.

\begin{definition}\label{def:supersafe}
    The \ith{i} head ($H_i$) of a \tdfa[2] $M$ is said to be \emph{supersafe}\footnote{This definition corresponds to a restricted version of the concept of \emph{safe} heads, introduced in~\cite{GS22}.} if and only if 
    there exists a \tdfa{} $N_i$ such that the following are true 
    for any input string $w$:
    \begin{enumerate}
        \item $N_i$'s head follows a finite  trajectory (\ie, a sequence of head position indices) $T_w$, 
        after which $N_i$ 
        halts, when given the input $w$. 
        \item If a (single\-/head) verifier simulates $M$ by using its own head to imitate the movements of $H_i$ on the input $w$ and receiving information about the other head from a prover, the trajectory of the verifier's head is guaranteed to be some prefix of $T_w$ 
        (after which the simulation ends by arriving at a halting state of $M$), regardless of whether the readings of the other head are provided consistently with the input string or are made up arbitrarily at every step by the prover.
    \end{enumerate}
\end{definition}

Intuitively, the transition function of such a \tdfa[2] renders the supersafe head's trajectory insensitive to the readings of the other head.

The runtime of a \tdfa[2] $M$ with a supersafe head must have a linear upper bound, since the \tdfa{} $N_i$ corresponding to the supersafe head is guaranteed to halt for all inputs, any \tdfa{} running on an input of length $n$ can attain $\OH{n}$ different configurations, and \Cref{def:supersafe} guarantees that $M$'s runtime is not longer than that of $N_i$.


As an example, consider the following \tdfa[2] for $\mixedeq{}$: The first head, which is supersafe, walks over the input from the left endmarker to the right, and then back towards the left. The second head moves one step to the right for every $a$ that the first head scans during its left\-/to\-/right phase, and attempts to move one step to the left for every $b$ scanned by the first head in its right\-/to\-/left phase. The machine rejects if the second head fails to return to the left endmarker or attempts to move beyond it in that last phase, and accepts otherwise.

 The class of languages that can be recognized by \tdfa[2]'s with at least one supersafe head will be denoted by $\TDFA[2_s]$.   It is easy to see that many other languages (\eg, \orderedeq, \powereq, \pal, and \twin) are also in $\TDFA[2_s]$.
We do not know if  $\TDFA[2]=\TDFA[2_s]$.

\begin{theorem}\label{thm:2dfass}
For any $\err>0$,     every language in $\TDFA[2_s]$ has an IPS with error bound $\err$ comprising of a constant\-/space (classical or quantum) prover and a  constant\-/space verifier that halts with probability at least $1-\err$ within expected polynomial time, and may fail to halt with the remaining probability.
\end{theorem}
\newcommand{\sslabel}{\ensuremath{\mathrm{ss}}}

\begin{proof}
Let $M$ be a \tdfa[2]  with a supersafe head $H_1$ and another head $H_2$, recognizing language $\langL$.  In relation to $M$ and $H_1$, let $N_1$ be the corresponding \tdfa{} described in \Cref{def:supersafe}.



Consider the  constant\-/space verifier $V$ described in \Cref{fig:supersafever}. The setting of the values of parameters $m$ and $p$ in that algorithm will be discussed later. The algorithm consists of $m$ rounds. In each round, the prover $P$ that communicates with $V$ is  supposed to report the readings of the head of $N_1$ at each step of $N_1$'s computation on the input string $w$, and $V$ randomly decides whether to use this information to  simulate $M$ or to check the correctness of $P$'s messages by simulating $N_1$. If, in any round, $V$ discovers that $P$ is lying about those readings, or a simulation of $M$ ends in rejection,  $V$ rejects. Otherwise, it accepts.

\begin{figure}[htb!]
    \caption{A verifier $V$ for a language recognized by \tdfa[2] $M$.}
    \label{fig:supersafever}
    \begin{turing}[CHOOSE]{V}{On input $w$:}
        \titem{Do the following for $m$ rounds:}
        \ttitem{Move $V$'s input head to the right endmarker and then return it to the left endmarker, thereby giving  $P$ time to move its own input head to the left endmarker.}
        \ttitem{Send $P$ the message \msg{start new round}.}
        \ttitem[CHOOSE]{Randomly pick one of the following simulations (\Cref{stg:SIMN1} or \Cref{stg:SIMM}) to execute in this round.  (In either case,  the prover will be asked to send the sequence of symbols that are supposed to be  scanned by the head of  $N_1$'s computation on input $w$ throughout that stage.)}
        \tttitem[SIM-$N_1$][stg:SIMN1]{
        \branchpr{1-p}
        Simulate $N_1$ on $w$, checking if the head readings match those reported by $P$ at every step. If a mismatch is detected, \reject{}. Otherwise, end this stage when the simulation reaches a halting state of $N_1$.}
        \tttitem[SIM-$M$][stg:SIMM]{
        \branchpr{p}
        Simulate $M$'s computation on $w$ by using the symbol sent by  $P$ as $H_1$'s reading at each step, and using $V$'s head to imitate $H_2$'s trajectory on the input.  If the simulation halts with rejection, \reject{}. If the simulation halts with acceptance, end this stage.}
        \ttitem{Send $P$ the message \msg{this round is over}.}
        \titem{(All rounds were completed without rejection.) \Accept.}
    \end{turing}
\end{figure}

$V$ does not trust $P$, which may lie about $H_1$'s readings in attempts to trick $V$ into accepting or even looping forever without reaching a halting state when the input is not in $\langL$.  $V$ will therefore rely on $P$'s messages with a very low probability $p \in \paren*{0, 0.5}$, and validate that  $P$ is indeed transmitting  $N_1$'s head readings with the remaining high probability, $1-p$.

\Cref{fig:honestprover} depicts an ``honest'' prover algorithm  that leads $V$ to acceptance with probability 1 whenever the input string $w$ is in $\langL$: No matter what sequence of random choices is made by  $V$, all rounds will be completed without rejection, since the prover is faithful to $N_1$, and simulations of $M$ based on its transmissions for the readings of $H_1$ end in acceptance by $M$. The runtime of $V$ in this case is polynomially bounded, since the simulations of both $N_1$ and $M$ will take linear time. 

\begin{figure}[htb!]
    \caption{A  prover that simulates $N_1$ on its input and reports its head readings in each round.}
    \label{fig:honestprover}
    \begin{turing}{P}{On input $w$:}
        \titem{Repeat the following in an infinite loop:}
        \ttitem{Move the input head to \lend.}
        \ttitem{Wait for $V$ to send the message \msg{start new round}.}
        \ttitem{Simulate $N_1$ on $w$, sending the head readings to $V$ at each step until $V$ sends the message \msg{this round is over}.}
    \end{turing}
\end{figure}

It remains to conclude the proof by showing that no prover algorithm $P^*$ can make $V$ fail to reject (\ie, cause it to accept or loop) with probability more than $\err>0$, where we can tune $\err$ to be as small as we desire by setting $m$ and $p$ appropriately, and also that the probability of $V$  halting in polynomial expected time can be tuned similarly. 

$V$ rejects any prover which does not provide a sequence of purported head readings of $N_1$ in response to its requests in \Cref{stg:SIMN1,stg:SIMM}.  At those stages, the symbols that $P^*$ provides will either match or  not match the actual readings of $N_1$.  In the following analysis, let $t_i$ denote the probability of the event that the symbols provided by $P^*$ match the readings of $N_1$ when $V$ is in its \ith{i} round of verification.

If the input $w$ is not a member of $\langL$, one of the following four combinations will occur in each round $i$ of $V$'s processing of $P^*$'s transmissions:
\begin{enumerate}
    \item $V$  simulates $N_1$, and determines that $P^*$'s transmissions about the head readings are correct.  In this case, $V$ will advance to the next round in $\OH{n}$ steps.  The probability of this event is $\paren*{1-p}t_i$.
    \item $V$  simulates $N_1$, and determines that $P^*$'s transmission about the head readings do not match what $V$ sees for itself.  In this case, $V$ will reject in $\OH{n}$ steps.  The probability of this event is $\paren*{1-p}\paren*{1-t_i}$.
    \item $V$ simulates $M$ while $P^*$ transmits the correct head readings for $N_1$.  In this case, $V$ will reject in $\OH{n}$ steps.  The probability of this event is $pt_i$.
    \item $V$ simulates $M$ while $P^*$ transmits fictitious head readings for $N_1$.  In this case, $V$ can reach an incorrect conclusion that $M$ accepts $w$ and pass this round, or it can be fooled into  ``simulating'' a nonexistent computation of $M$ forever.  The probability of this event is at most $p\paren*{1-t_i}$.
\end{enumerate}
Based on this, the probability that $V$ will reject, i.e., $\pr_{\rej}(V,P^*,w)$, is at least
    $f(1)$,
where the function $f$ is defined recursively as follows:
\begin{equation*}
    f(i) = \begin{cases}(1-p)t_if(i+1) + (1-p)(1-t_i) + pt_i & \text{for } i < m\\
    (1-p)(1-t_m) + pt_m & \text{for } i = m\end{cases}
\end{equation*}

Consider the following series of observations:
\begin{observations}
    \item $\paramset[\Big]{f(i)}_{t_i=0} = 1-p$.
    \item 
   $\paramset[\Big]{f(i)}_{t_i=1} = \begin{cases}f(i+1)+p\paren*{1-f(i+1)} & \text{for } i < m\\p & \text{for } i = m.\end{cases}$
    \item\label{obs:filowerbound} $f(i)$ is linear in terms of $t_i$, and $t_i$ is between $0$ and $1$. Thus, $f(i)$ is between the above two extremes.  This implies  that 
     \begin{equation*}
         f(i) \ge \begin{cases}\min\Set{1-p, f(i+1)+p\paren*{1-f(i+1)}} & \text{for } i < m\\
         \min\Set{1-p, p} & \text{for } i = m.
         \end{cases}
     \end{equation*}
    \item \Cref{obs:filowerbound} implies that if some $f(i)$ is greater than or equal to $1-p$, then so are all $f(j)$ for $j < i$, and in particular, $f(1) \ge 1-p$.
    \item Finally, for all $p > 0$, there exists an $m$ that forces there to exist an $f(i) \ge 1 - p$.  To prove this by contradiction, assume that there exists a $p > 0$ such that for all $m$, $f(i) < 1 - p$ (equivalently, $1-f(i)>p$) holds for all $f(i)$.
    However, then, for such a $p$ and in conjunction with \Cref{obs:filowerbound}, we get
    \begin{equation*}
        1 - p > f(i) \ge f(i + 1) + p(1 - f(i + 1))
    \end{equation*}
    for $i < m$. Applying this inequality repeatedly, and then using the assumption once again, we get
    \begin{equation*}
        f(1) \ge f(m) + \sum_{i=2}^m p(1 - f(i)) > f(m) + (m-1)p^2.
    \end{equation*}
    Regardless of what $p>0$ is, a large enough $m$ (say, $2\ceil*{\sfrac{1}{p^2}} + 1$) results in $f(1) > 1$, a contradiction.
\end{observations}
We therefore conclude that the error bound $\err$ for $V$ failing to reject with $P^*$ can be brought arbitrarily close to 0: For any desired value of $\err$, setting $p = \err$ and fixing $m$ to an accordingly large value that exists by the argument above guarantees that $\pr_{\rej}(V,P^*,w)\geq 1-\err$.

The probability that $V$ can be tricked to loop is also bounded by $\err$. With the remaining probability, $V$'s runtime will be bounded by $\OH{n}$.
\end{proof}

\section{Proofs of quantumness: A definition}\label{sec:PoQdef}

We now present a framework where a classical verifier in an interactive proof system can distinguish whether the prover it is communicating with is a quantum machine or not. Although we will see that one can presently  obtain ``unconditional'' conclusions about quantumness for very slowly growing space bounds,  our definition also accommodates  similar protocols between bigger machines by allowing the verifier and the prover larger space bounds.

In the following, we consider a space\-/bounded verifier (\Cref{def:BUSECVERIFIERDEF}) which halts with high probability within polynomial expected runtime, and which can be paired by any quantum or classical prover operating under the same space bound to form an interactive proof system. 
\begin{definition}\label{def:PoQ}
 A  verifier $V$ is said to \emph{check proofs of quantumness (with halting probability $p_h$)} if 
\begin{itemize}
    \item there exists a number $p_h > \sfrac{1}{2}$ such that, on any input, $V$ halts with probability at least $p_h$ within expected polynomial time, and fails to halt with the remaining probability, 
    \item $V$ runs within space $s(n)$ for some function $s$ of the input length $n$, and
    \item there exists 
a number (the \emph{success gap}) $\Delta>0$ and a quantum prover $P^Q$ that runs within space $s(n)$ such that, for every classical prover $P^C$ that runs within space $s(n)$, there exists an infinite set of strings $W_{P^C}$ such that 
\begin{equation*}
    \inf \Set{ \pr_{\acc}\paren*{V,P^Q,w} | w \in W_{P^C} } - \sup \Set{ 1-\pr_{\rej}\paren*{V,P^C,w} | w \in W_{P^C} } \geq \Delta.
\end{equation*}
\end{itemize}
\end{definition}

Intuitively, $V$ expects the prover  to prove that it can solve a problem instance encoded in their common input string $w$ correctly. As we will see below, one can start the construction of such a protocol by picking the underlying computational problem to be one that is provably impossible for an $\oh{\log \log n}$\-/space \ptm{}, but feasible for a \tqcfa{}. This will mean that there exist infinitely many input strings (set $W_{P^C}$ in \Cref{def:PoQ}) for which a particular classical prover $P^C$ claiming to be able to solve the problem will have to ``make up'' an answer, which will be incorrect, (and whose purported proof will therefore be ``caught'' as erroneous) by $V$ with a significant probability. On the other hand,  the quantum prover employing the correct algorithm to solve the problem is also supposed to prove its result to $V$. We will note that not all of the problems with quantum advantage listed in \Cref{subs:pfaqfa} seem to be associated with such ``proofs'' that can be generated and communicated between  machines with so small resource bounds. For the problems which do accommodate such proofs,  $V$ will accept the claim (to quantumness) of a genuine quantum prover running the correct algorithm with considerably higher probability than that of any small\-/space \ptm{} $P^C$ with the same claim when running on strings  in $W_{P^C}$.\footnote{The ``parity game''~\cite{AB09,CHSH69}, which cannot be won by any purely classical couple with probability greater than $0.75$, whereas  a couple making use of quantum entanglement is ensured to win with probability greater than $0.85$, can be seen as another ``test setup'' where the distinction is based on such a gap between success probabilities.}

Let us take a moment to consider the ordering of the quantifiers in \Cref{def:PoQ}. At first sight,  one may wonder whether we can use the same set of strings to distinguish between the ``correct'' (quantum) prover (which is supposed to succeed on all members of all such ``test sets'' anyway) and all classical provers, and move the quantifier for the input strings to the left accordingly. This is, however, impossible for the following reason: We will be presenting many example protocols where, as mentioned above, the prover is supposed to first solve a problem that is difficult for small\-/space \ptm{}'s and then generate and transmit a proof of correctness of the obtained answer. In those protocols, once  an answer to the original problem is at hand, this second stage of communicating the proof is an ``easy'' task (for even a \tpfa{}) to realize. Therefore, if the test set (say, $W$) is fixed, then for any one of its finite subsets, say, $W'$, there exists   a classical prover $P'$ which is ``hardwired'' to respond to inputs in $W'$ by communicating the correct answers and their proofs to the verifier. Such a prover would succeed with very high probability on members of $W'$, and so the success gap required by the inequality at the end of the definition would not be achieved.

We would like \Cref{def:PoQ}  to represent a realistic framework for testing some device for quantumness in a reasonable amount of time. This necessitates us to clarify an aspect about the runtimes of the provers appearing in the interactive proof protocols to be described below. In most of the early work on interactive proof systems, the prover was modeled as a magically powerful entity which could compute anything (including uncomputable functions) in a single step. The limitations that later papers~\cite{conlad95,GKR15} did impose on the prover are different than ours in an important respect: Condon and Ladner~\cite{conlad95} study provers that are restricted to have polynomial\-/size strategies, but this still means that any actual machine in the role of the prover would need superpolynomial runtime under standard complexity assumptions. Goldwasser et al.~\cite{GKR15} do impose a polynomial time bound on the prover, but their model clearly does not require the prover to run simultaneously with the verifier, whose runtime is significantly shorter. In our model, the understanding is that the prover and the verifier  start working simultaneously, at which point they access the input tape for the first time, and the process ends when the verifier halts by accepting or rejecting. In our examples, the verifier will be imposing a polynomial time limit on the prover to declare its answer to the problem instance, so the prover will not be able to run an exponential\-/time algorithm to  completion for many inputs. On the other hand, the small\-/space restriction introduces a complication that verifiers with $\Omega(\log n)$ memory do not face: Our protocols will allow  some malicious provers to fool the corresponding verifier, which is not able to measure the runtime while simultaneously checking a purported proof, to run forever without reaching a verdict, with nonzero probability. This is why \Cref{def:PoQ} requires a gap between the minimum acceptance probability caused by a quantum prover and the maximum probability that any classical prover can trick the verifier either to  announce acceptance or to run forever instead of rejecting.

We will use the following common template when constructing verifiers that check proofs of quantumness: One starts with a promise problem $\mathbf{P}=\paren*{\mathbf{P}_{\Yes}, \mathbf{P}_{\No}}$ that is  impossible for small\-/space polynomial\-/time probabilistic machines, but computable by quantum machines within those resource bounds. $\mathbf{P}$ should also have the property that each question's correct answer has an interactive proof where both sides are small\-/space  machines. We build a protocol where a verifier first runs a polynomial\-/time ``clock'' to give the prover time to compute and announce the answer to the common input string. The prover is then supposed to prove the correctness of this answer to the verifier with high probability in polynomial time. We say that a verifier built according to this template \emph{checks proofs of quantumness based on $\mathbf{P}$}.

\section{Proofs of quantumness based on \dsfk{} problems}\label{sec:padded}

In this section, we will demonstrate a method to convert some of the exponential\-/time \tqcfa{} algorithms for the languages in \Cref{tab:languagesrecognizedinexp} to (polynomial\-/time) tests of quantumness. An important tool to be used in that regard is the following theorem proved by Freivalds and Karpinski~\cite{FK94}, building on 
work by Dwork and Stockmeyer~\cite{DS92}.

\begin{theorem}\label{thm:FK}
Let $\langA, \langB \subseteq \Sigma^*$ with $\langA \cap \langB = \emptyset$. Suppose there is an infinite set $I$ of positive integers and functions $g(m), h(m)$ such that $g(m)$ is a fixed polynomial in $m$, and for each $m \in I$, there is a set $W_m$ of words in $\Sigma^*$ such that:
\begin{enumerate}
    \item $\abs{w} \leq g(m)$ for all $w \in W_m$,
    \item there is a constant $c > 1$ 
    such that $\abs{W_m} \geq c^m$, 
    \item for  every $w, w' \in W_m$ with $w \neq w'$, there are words $u, v \in \Sigma^*$ such that:
    \begin{enumerate}
        \item $\abs{uwv} \leq h(m), \abs{uw'v} \leq h(m)$, and
        \item either
            $uwv \in \langA$ and $uw'v \in \langB$, or 
            $uwv \in \langB$ and $uw'v \in \langA$.
    \end{enumerate}
\end{enumerate}
Then, if a \ptm{} with space bound $s(n)$ separates $\langA$ and $\langB$, then $s(h(m))$ cannot be in $\oh{\log m}$.
\end{theorem}

\begin{definition}\label{def:padded}
    For any language $\langL \subseteq \Sigma^*$, the padded language $\padlang{\langL} \subseteq \Sigma_\padsymb^*$ is obtained by appending the corresponding string $\padsymb^{2^{\abs{x}}-\abs{x}}$ 
    to each string $x \in \langL$, where $\Sigma_\padsymb = \Sigma \cup \Set{\padsymb}$ for some $\padsymb \notin \Sigma$.  Formally,
    \begin{equation*}
        \padlang{\langL} = \Set{ xy | x \in \langL, y = \padsymb^{2^{\abs{x}}-\abs{x}} }.
    \end{equation*}
\end{definition}

Given a string $xy \in \padlang{\langL}$, we refer to the prefix $x \in \langL$ as the \emph{core} (of $xy$) and the suffix $y \in \Set{\padsymb}^*$ as the \emph{padding} (of $xy$).

We note that it is impossible for a \ptm{}  running in space $\oh{\log \log n}$ to determine whether its input is in this form with the relationship prescribed in \Cref{def:padded} between the length of the core and the padding in polynomial time: The set of strings of the form $xy$, where $x \in \Sigma^*$ and $y = \padsymb^{2^{\abs{x}}-\abs{x}}$ is nonregular, and no such small\-/space machine can recognize a nonregular language in polynomial expected time~\cite{DS90}.\footnote{In the \nameref{sec:generalizedDS}, 
we generalize the result by Dwork and Stockmeyer~\cite{DS90} on the inability of these machines to recognize regular languages in polynomial expected time, showing that this remains true for \ptm{}'s that do not necessarily halt with probability $1$, as long as the ``halting part'' of the computation has polynomial expected runtime.
}
(We do not know if this task of checking for an exponential\-/length padding in the input is also impossible for small\-/space polynomial\-/time \qtm{}'s.\footnote{Note that the \power{} language (\Cref{tab:languagesrecognizedinexp}) can be recognized by an exponential\-/time \tqcfa{}.})

Consider the languages \padlang{\pal} and \padlang{\setneg{\pal}}, where \pal{} is the set of palindromes on the alphabet $\Sigma=\Set{a,b}$.
In the template of \Cref{thm:FK}, let $I$ be a set containing only even numbers, $g(m)=m,  h(m)=2^m$, and $W_m=\Sigma^{\sfrac{m}{2}}$. For every $w, w' \in W_m$, let
the corresponding $u$ be the empty string, and the corresponding $v$ be the string $w^R  \padsymb^{2^{m}-m}$. It follows that  no \ptm{} with a space bound in $\oh{\log \log n}$ can separate  \padlang{\pal} and \padlang{\setneg{\pal}}.

\pal{} is just one of a number of interesting languages that lead to this kind of impossibility result. Call any separation problem $(\langA,\langB)$ that matches the pattern specified in \Cref{thm:FK} with the functions $g$ and $h$ respectively set to be $m \mapsto m$ and $m \mapsto 2^m$ (as in the above example) a \emph{\dsfk{} problem}.\footnote{From the names of Dwork, Stockmeyer~\cite{DS92}, Freivalds, and Karpinski~\cite{FK94}.} All \dsfk{} problems are impossible for $\oh{\log \log n}$\-/space \ptm{}'s. In the following, we note a few more \dsfk{} problems based on languages we saw in \Cref{tab:languagesrecognizedinexp}. $\estring$ denotes the empty string.

To see that \padpair*{\twin} is a \dsfk{} problem, set  $I$ to be the set of all  odd numbers greater than 1, and $W_m=\Set{a,b}^{\frac{m-1}{2}}$. For any pair of distinct elements $w, w' \in W_m$, let $u=\estring$, and  $v=\#w\padsymb^{2^m-m}$.

To see that \padpair*{\mult} is a \dsfk{} problem,  set  $I$ to be the set of all  odd numbers greater than 5, and $W_m=\Set{1}\Set{0,1}^{\frac{m-5}{2}}$. For any pair of distinct elements $w, w' \in W_m$, let $u=\estring$, and  $v=\#1\#w\padsymb^{2^{m}-m}$.

It is easy to see that \padpair*{\mixedeq} is \emph{not} a \dsfk{} problem, by considering the following \tpfa{} $M$ that performs the separation: On an input of the form $xy$, where $x$ is the core, $M$ simply runs Freivalds' \tpfa{} algorithm for recognizing $\mixedeq{}$~\cite{F81} by treating $x$ as the sole input. The resulting algorithm's runtime is exponential in $\abs{x}$, but only polynomial in the overall input length. 

We are now ready to present our first proof\-/of\-/quantumness protocol.

\begin{theorem}\label{thm:paddedprotocol}
    Let $\langL$ be any language in $\BQTISP{t(n),s(n)} \cap \TDFA[2_s]$ such that $t(n)=2^{\OH{n}}$, $s(n)=\oh{\log \log n}$, and \padpair*{\langL} is a \dsfk{} problem.
    For any number $p_h<1$, there exists a verifier $V_{\langL}$ that checks proofs of quantumness based on \padpair*{\langL} with halting probability $p_h$.    
\end{theorem}
\begin{proof}
    For any language $\langL$ with the properties described in the theorem statement, let $Q_{\langL}$ be a \qtm{} that recognizes $\langL$ with error bound $\err_{Q_{\langL}}$  within space $s(n)=\oh{\log \log n}$ and expected runtime $2^{k n}$, 
    for  some integer $k>0$ and all sufficiently large input lengths $n$. Let $M$ and $M'$ be the \tdfa[2]'s (with at least one supersafe head) that recognize $\langL$ and $\setneg{\langL}$, respectively.\footnote{$\TDFA[2_s]$ is closed under complementation.} We describe a verifier $V_{\langL}$ (\Cref{fig:asdver}) and quantum prover $P^{Q}_{\langL}$ (\Cref{fig:asdprov}) that will be shown to lead $V_{\langL}$ to acceptance with high probability for every input satisfying the promise. 
    $V_{\langL}$ has parameters named $k$, $c_1$, $\errprematuretimeout$, and $\err_V$ that will be set to appropriate values, as described below. Since $P^{Q}_{\langL}$ runs only on the very short core $x$ (of length $m = \log n$) of the input string, its expected runtime will be short enough so that it will be able to run to completion before the ``clock'' stage of  $V_{\langL}$ terminates. After the prover announces its answer to the separation problem, it is supposed to  engage in a dialogue with $V_{\langL}$ to prove that the appropriate \tdfa[2] ($M$ or $M'$, depending on the announced answer) really accepts the input $x$. We now provide a detailed analysis.

\begin{figure}[htb!]
    \caption{A verifier based on a \dsfk{} problem \padpair*{\langL}, where $\langL\in \TDFA[2_s]$.}
    \label{fig:asdver}
    \begin{turing}[VERIFY]{V_{\langL}}{On input $xy$ of length $n$, such that $x \in \Sigma^*$, $\abs{x}=\log_2 n$, and $y \in \padsymb^*$ for $\padsymb \notin \Sigma$:}
        \titem[CLOCK][stg:CLOCK1]{Run a ``clock'' (\Cref{fact:factclock}) which has an expected runtime in $O(n^{k+1})$, and terminates in fewer than $c_1 n^k$ steps with probability at most $\errprematuretimeout$. At every step of this procedure, ask the prover if it has computed a \Yes{}/\No{} answer for $xy$, and pass to the next stage if you receive such an answer. \Reject{} if the prover has still not sent a claim about $xy$ when the clock times out.}
        \titem[VERIFY][stg:VERIFY]{If the prover has answered \msg{\Yes} (resp.\ \msg{\No}), 
        run the verification algorithm specified in \Cref{fig:supersafever} (codified to treat only the $x$ segment on the tape as its input string) to check the  membership of $x$ in $\langL$ (resp.\ $\setneg{\langL}$) with error bound $\err_V$ using the \tdfa[2] $M$ (resp.\ $M'$) described in the text. \Accept{} (resp.\ \reject) if that algorithm accepts (resp.\ rejects).}
    \end{turing}
\end{figure}

\begin{figure}[htb!]
    \caption{A quantum prover based on \padpair*{\langL}, where $\langL\in \BQTISP{2^{kn},s(n)}\cap \TDFA[2_s]$.}
    \label{fig:asdprov}
    \begin{turing}{P^{Q}_{\langL}}{On input $xy$ of length $n$, such that $x \in \Sigma^*$, $\abs{x}=\log_2 n$, and $y \in \padsymb^*$ for $\padsymb \notin \Sigma$:}
        \titem{Run the \qtm{} $Q_{\langL}$ (codified to treat only the $x$ segment on the tape as its input string) to decide whether $x$ is in $\langL$ or not. At every step during this procedure, give the answer \msg{still computing} to each question of the verifier. When $Q_{\langL}$ halts, send its result (\msg{\Yes} or \msg{\No}) to the verifier.}
        \titem{Run the ``proof'' algorithm specified in \Cref{fig:honestprover} (codified to treat only the $x$ segment on the tape as its input string) using the \tdfa{} associated with the supersafe head of the  machine $M$ (or, identically, $M'$) described in the text.} 
    \end{turing}
\end{figure}

Consider the behavior of $V_{\langL}$ when faced with the prover $P^{Q}_{\langL}$ on some common input string $xy$ on the input alphabet $\Sigma_\padsymb$.  $P^Q_{\langL}$'s expected runtime on sufficiently long inputs is at most $2^{km} = 2^{k \log n}= n^k$.
With probability at least $1 - \errprematuretimeout$, $V_{\langL}$ will  grant $P^Q_{\langL}$ at least $c_1 \cdot n^k$ steps to announce an answer. 
By Markov's inequality, $Q_{\langL}$'s  computation will need more than $c_1 \cdot n^{k}$ steps with probability at most $\sfrac{n^k} {c_1 n^k}=\sfrac{1}{c_1}$. 
If it can run to completion, $Q_{\langL}$ will return the correct answer with probability at least $1-\err_{Q_{\langL}}$.
Since the VERIFY stage of $V_{\langL}$ will accept proofs of a correct answer with probability 1, 
we conclude that $\inf \Set{\pr_{\acc}\paren*{V_{\langL},P^{Q}_{\langL},w} | w \in 
\Sigma^m\Sigma_\padsymb^{2^m-m}}$ nears at least
\begin{equation}
    \paren*{1-\errprematuretimeout} \cdot \paren*{1 - \sfrac{1}{c_1}} \cdot \paren*{1 - \verr_{Q_{\langL}}}\label[expression]{eqn:ifade}    
\end{equation}
as $n$ (and therefore $m$) grows. 
Recall from \Cref{subs:pfaqfa} that there exists a $Q_{\langL}$ for any desired small positive value of $\verr_{Q_{\langL}}$, and a choice of $\verr_{Q_{\langL}}$ sets $k$ to a corresponding value. Note that  $\errprematuretimeout$ 
can be set from the outset to a positive value that is as small as desired by plugging in an appropriate parameter for the clock, 
and $c_1$ can be selected to be as large as necessary. The probability  (\Cref{eqn:ifade})  that $V_{\langL}$ will accept a proper quantum prover can therefore be set to be as near to $1$ as desired. $P^{Q}_{\langL}$ runs within space $\oh{\log \log n}$.\footnote{Note that $P^{Q}_{\langL}$ would still run within space $\oh{\log \log n}$ on inputs satisfying the promise of having exponentially long padding  if we allowed $Q_{\langL}$ a  bigger space budget of $s(n)=\oh{\log n}$, but \Cref{def:PoQ} stipulates that the prover should respect the common space bound on all possible inputs.}


Let us now analyze what happens when $V_{\langL}$ is faced with some classical $s(n)$\=/space prover $P^C$.
If $P^C$ fails to announce a \Yes{}/\No{} answer during the CLOCK stage of $V_{\langL}$, it will be rejected by probability $1$.
Since no  \oh{\log \log n}\=/space \ptm{} can separate $\padlang{\langL}$ and $\padlang{\setneg{\langL}}$, any algorithm realized by $P^C$ to return an answer during the clock stage must answer incorrectly (with probability at least $\frac{1}{2}$) for an infinite\footnote{Since a machine that performs the separation answers all inputs correctly with probability greater than $\frac{1}{2}$,  any machine $T$ that \emph{fails} to separate $\padlang{\langL}$ and $\padlang{\setneg{\langL}}$ must be associated with  a nonempty set of strings that are answered correctly  with probability at most $\frac{1}{2}$. Furthermore, this set must be infinite, since any \ptm{} which acts undesirably in response to only finitely many input strings can be ``repaired''  without changing its time or space bounds.} set of input strings $W'_{P^C}$.
This incorrect answer will be exposed (with probability at least $1-\verr_V$, by \Cref{thm:2dfass}) when it is checked against the judgment of $M$ (or $M'$) in the VERIFY stage, leading to $P^C$ being rejected with probability at least $\sfrac{1-\verr_V}{2}$ in this case.
$W'_{P^C}$ has an infinite subset $W_{P^C}$ of strings that are sufficiently long for the above\-/mentioned lower bound for the acceptance probability associated by $P^{Q}_{\langL}$ to hold.
We conclude that, for any such classical prover $P^C$, 
\[\sup \Set{ 1-\pr_{\rej}\paren*{V_{\langL},P^C,w} | 
w \in W_{P^C} }\leq \sfrac{1+\err_V}{2}.\]
$\err_V$, which also bounds the probability that $V_{\langL}$ may be tricked to loop forever, is a value that can be set to be as small as one desires too, and therefore the success gap
\begin{equation*}
    \inf \Set{ \pr_{\acc}\paren*{V_{\langL},P^{Q}_{\langL},w} | w \in W_{P^C} } - \sup \Set{ 1-\pr_{\rej}\paren*{V_{\langL},P^C,w} |  w \in W_{P^C}} 
\end{equation*}
can be set to be greater than any desired value $\Delta<\frac{1}{2}$.

As established in \Cref{thm:2dfass}, the runtime of the VERIFY stage is linear in the length of the core (except in the low\-/probability case where $V_{\langL}$ is deceived into looping), so  $V_{\langL}$ halts with probability $1-\err_V$ in expected time $\OH{n^{k+1}}$.
\end{proof}

Since \pal{} and \twin{} are known to satisfy all the requirements of \Cref{thm:paddedprotocol}, the associated problems \padpair*{\pal} and \padpair*{\twin} are suitable for testing claims to quantumness by small\-/space machines without relying on unproven assumptions about the impossibility of a particular problem for a specific model of computation.

\section{Verification beyond \texorpdfstring{\TDFA[2]}{2DFA(2)}}\label{sec:squ}

The verifier template described in the proof of \Cref{thm:paddedprotocol} can be modified to support protocols based on problems with different properties. 
Consider the language \squarelang{}, which was described in \Cref{tab:languagesrecognizedinexp}. For any error bound $\err_Q>0$, this language is recognized by a \tqcfa{} whose expected runtime on all inputs of sufficient length $n$ is at most $2^{k n}$ for some  $k>0$ corresponding to $\err_Q$. It is not known whether \squarelang{} is  recognizable by either a \tdfa[2] or a sublogarithmic\-/space \ptm{}. Let  $\subsquare = \Set{ a^i b^j | j <  i^2}$. We will describe a protocol whereby a quantum prover can demonstrate its ability to solve the separation problem $(\padlang{\squarelang}, \padlang{\subsquare})$ to a classical  verifier $V$, depicted in \Cref{fig:sqver}. 

The parameters $k$, $\errprematuretimeout$, $c_1$, $c_F$, $d_F$, $p$, and $h_V$ in the description of $V$ allow us to tune the success gap of the protocol, as will be explained below. On an input string of the form $a^i b^j \padsymb^*$, this verifier expects a  prover to first announce its claim about the input string's membership in the allotted time, and then to repeatedly  transmit segments composed of $i$ symbols. One probabilistic branch of  $V$ verifies that each segment (in a sufficiently long prefix of this transmission) is indeed of the required length, while another branch uses this ``ruler'' provided by the prover to determine whether the block of $b$'s in the input consists of $i$ mini\-/blocks of length $i$, i.e., has length $i^2$. For this purpose,  $V$ compares the number of segment separator symbols transmitted by the prover in the time it takes $V$ to scan the segment of $b$'s in the input with the number of $a$'s in the input, using the technique invented by Freivalds~\cite{F81} to build a \tpfa{} that recognizes $\orderedeq{}$.   A detailed analysis will follow the description of the quantum prover below.  

\begin{figure}[htb!]
    \caption{A verifier $V$ based on $(\padlang{\squarelang}, \padlang{\subsquare})$}
    \label{fig:sqver} 
    \begin{turing}[CHECK-RULER]{V}{On input $xy$ of length $n$, such that $x = a^ib^j$, $j \leq i^2$, $\abs{x}=\log_2 n$, and $y \in \padsymb^*$:}
        \titem[CLOCK][stg:CLOCK2]{
            Run a ``clock'' (\Cref{fact:factclock}) which has an expected runtime in $O(n^{k+1})$, and terminates in fewer than $c_1 n^k$ steps with probability at most $\errprematuretimeout$. At every step of this procedure, ask the prover ($P$) if it has computed a \Yes{}/\No{} answer for $xy$, and pass to the next stage if you receive such an answer. \Reject{} if the prover has still not sent a claim about $xy$ when the clock times out.}
        \titem{
            Move the input head to the right end of the core $x$ and then return it to the left endmarker, thereby giving  $P$ time to move its own input head to the left endmarker.}
        \titem{
            Send $P$ the message \msg{start proof}.}
        \titem[CHOOSE][stg:CHOOSE2]{
            Randomly pick one of the following two procedures (CHECK-RULER or FREIVALDS) to execute.  (In either case, the prover will be expected to transmit a  sequence of ``$z^{i-1} \#$'' segments  throughout that stage.  $V$ will drive this transmission by requesting each symbol, pausing for one step after each $\#$ to give the prover (see \Cref{fig:sqpro}) time to move its head to the proper location for starting the  next segment.)}
        \ttitem[CHECK-RULER][stg:A]{\branchpr{p}
            Repeatedly move the head between the two ends of the maximal prefix of $a$'s in the input. On each $a$, throw a coin with a probability $2^{-h_V}$ of coming up heads. If that coin comes up heads in all $i$ steps of a walk of the head from one end of the $a^i$ segment to the other, \accept{}.  If the prover is detected to fail to transmit segments of exactly $i-1$ contiguous $z$'s punctuated by $\#$'s, \reject{}.}
        \ttitem[FREIVALDS][stg:B]{\branchpr{1-p}
            Initialize counters $C_a$ and $C_\#$ to 0, and then repeatedly move the input head between the two ends of the core $a^i b^j$. In each pass of the input,  perform the following controls parallelly:}
        \tttitem{
            Count both the number of $a$'s in the input and the number of $\#$'s sent by $P$ while the input head is 
            scanning $b$'s modulo the parameter $c_F$.  
            If the two counts are not congruent (mod $c_F$) at the end of the pass, or if the prover does not send a $\#$ to coincide with the last $b$ that is scanned in this pass, conclude that $j\neq i^2$ and \accept{} (resp.\  \reject{}) if the answer previously announced by $P$ agrees (resp.\ disagrees) with this conclusion.}
        \tttitem{
            Throw a fair coin for each $a$ that the input head scans in the input (call these ``the group of $a$ coins'') and for each $\#$ sent by $P$ while the input head is scanning a $b$ (``the $\#$ coins''). If \emph{all} coins in one of these groups turn up heads during a complete (left\-/to\-/right or right\-/to\-/left) sweep of the core by the input head, this is a ``goal'' scored by that group. If exactly one group scores a goal during a sweep, this is a ``win'' for that group. If this sweep is a win for the group of $a$ (resp.\ $\#$) coins, increment the counter $C_a$ (resp.\ $C_\#$) by 1. If the total number of wins reaches $d_F$, terminate the loop and check the counter values. If both counter values are nonzero, \accept{} (resp.\ \reject{}) if the answer sent by $P$ in the CLOCK stage was \msg{\Yes} (resp.\ \msg{\No}). If a counter value is zero,  \reject{} (resp.\ \accept{}) if the answer sent by $P$ in the CLOCK stage was \msg{\Yes} (resp.\ \msg{\No}).}
    \end{turing}
\end{figure}

\begin{figure}[htb!]
    \caption{A quantum prover based on $(\padlang{\squarelang},\padlang{\subsquare})$}
    \label{fig:sqpro}
    \begin{turing}{P^{Q}}{
    On input $xy$, where $x = a^ib^j$, $y \in \padsymb^*$, and $\abs{x}= \log_2 \abs{xy}$:}
        \titem{Run the \tqcfa{} $Q_{\squarelang}$ (codified to treat only the $x$ segment on the tape as its input string) to decide whether $x\in \squarelang$ or not. At every step during this procedure, give the answer \msg{still computing} to each question of the verifier. When $Q_{\squarelang}$ halts, send its result (\msg{\Yes} or \msg{\No}) to the verifier.}
        \titem{Move the input head to the first $a$ in the input.} 
        \titem{Wait for $V$ to send the message \msg{start proof}.}
        \titem{Repeatedly move the head between the two ends of the block of $a$'s on the input tape. 
        On each pass, perform the following actions:}
        \ttitem{Skip the first $a$. For all the remaining $a$'s, send a $z$ to the verifier upon its symbol request when the head is on this $a$. For the non\-/$a$ symbol that you scan in the input tape when you walk off the block, send a $\#$ to the verifier upon its symbol request.}
    \end{turing}
\end{figure}

The first stage of the quantum prover $P^{Q}$ in \Cref{fig:sqpro} runs the  \tqcfa{} algorithm $Q_{\squarelang}$ for recognizing \squarelang{} on the core of the form $a^i b^j$ in the same fashion as the prover in \Cref{fig:asdprov}.  Since the CLOCK stage of the verifier in \Cref{fig:sqver} is identical with that of \Cref{fig:asdver},  the first stage of $P^{Q}$ will have sufficient time to finish $Q_{\squarelang}$ and report a correct \Yes{}/\No{} answer with a probability that can be set to be as close to 1 as one desires by choosing  an appropriately low\-/error version of $Q_{\squarelang}$, and plugging in the values of $k$, $c_1$, and \errprematuretimeout{} 
 accordingly, by the analysis we presented in the proof of \Cref{thm:paddedprotocol}. In the remainder of its execution, the prover responds to the symbol requests from the verifier by sending an infinitely long  sequence $z^{i-1}\#z^{i-1}\#z^{i-1}\#\dotsm$ made up of segments of length $i$. Note that the ``proof'' transmitted by $P^{Q}$ does not depend on the answer that it announces earlier.

In the CHOOSE stage of \Cref{fig:sqver}, $V$ will choose to execute the control in the CHECK-RULER stage with probability $p$. Since $P^{Q}$ transmits the proper symbol sequence dictated by the protocol, $V$'s CHECK-RULER stage will accept with probability 1 within expected runtime $i 2^{h_V i}$ when faced with $P^{Q}$. 
If $V$ chooses to execute the FREIVALDS stage,\footnote{This happens with  probability  $1-p$.} $V$ simply runs a modified version of Freivalds' algorithm for $\orderedeq{}$, which has been codified to compare the number of $a$'s in the core with the number of $\#$'s transmitted by $P^{Q}$ while $V$'s input head is on the $b$\=/block of the core. 
Since $P^{Q}$ is honest to its mission, it will transmit the same number of $\#$'s during every sweep of the core. As proven by Freivalds~\cite{F81}, one can set the parameters $c_F$ and $d_F$ to guarantee that the FREIVALDS stage arrives at the correct decision about whether the two numbers being compared are equal or not with probability $1-\err_F$, for any desired value of $\err_F>0$. 
That stage rejects $P^{Q}$ only if $V$'s conclusion about the core disagrees with the answer previously sent by $P^{Q}$. We have already established that we can make that answer's correctness probability as close to 1 as we wish, and 
the parameters  
$c_F$ and $d_F$ can be set to tune the probability of a disagreement with that correct answer to a value below any desired positive bound as long as the correct ruler is transmitted. 
We conclude that  $\inf \Set{ \pr_{\acc}\paren*{V,P^{Q},w} | w \in W }$ can be arranged to be arbitrarily close to 1 for an infinite set $W$ containing all padded strings that are sufficiently long.

Note that, if it chooses to execute the FREIVALDS stage, $V$ bases its decision  on the assumption that the prover $P$ that it is communicating with will send a proper sequence of segments, each of length $i$. The runtime of Freivalds' algorithm, which $V$ executes in this case, supplies a bound on the number of symbols that $V$ will request from $P$.  That algorithm's expected runtime is $\OH{m 2^r}$, where $m$ is the length of the string it is working on, $r$ is the number of coins it tosses during a single pass of that string, and the parameters $c_F$ and $d_F$ determine the constant of proportionality ``hidden'' by the big\=/\justOH{} notation~\cite{DS90}. In this case, the promise of the problem guarantees that  $m\leq i+i^2$ and $r\leq 2i$. We can thefore say that, for sufficiently long inputs, this procedure has expected runtime under $2^{k_F i}$, for a sufficiently large number $k_F$ that depends on the desired $\varepsilon_F$.
Consider a parameter $K>1$. The probability of the runtime of the FREIVALDS stage exceeding $K2^{k_F i}$ is at most $\frac{1}{K}$, by Markov's inequality. 
Since $i>0$,  $\frac{1}{K}$ is also an upper bound for the probability that $V$ will request more than $iK2^{k_F i}$ symbols from the prover during this stage.

Let us now consider a classical prover $P^{C}$ interacting with $V$. Note that $V$ will reject any prover that fails to return a \Yes{}/\No{} answer, or to respond to a symbol request promptly, with probability $1$. Assume that $P^{C}$ has reported a \Yes{}/\No{} answer within the time allotted by the CLOCK stage of $V$. We will analyze $P^{C}$'s chances of being accepted in the cases where it adopts one of the following strategies about which symbols to transmit in response to the verifier's subsequent requests:
\begin{strategies}
    \item\label{strat:1} Transmit a sequence which fails to match the ``expected'' pattern $(z^{i-1}\#)^+$ in at least one position among the first $iK2^{k_F i}$ symbols
    \item\label{strat:2} Transmit a sequence whose first $iK2^{k_F i}$ symbols  match the ``expected'' pattern $(z^{i-1}\#)^+$ 
\end{strategies}

If $P^{C}$ employs \Cref{strat:1} and $V$ chooses (with probability $p$) to execute the CHECK-RULER stage, $V$ will reject the prover, unless that stage terminates before $P^{C}$ attempts to transmit the first ``defect'' in the pattern. In this stage, $V$ repeatedly sweeps the block of $a$'s in the input from end to end, and decides to halt and accept with probability $2^{-h_V i}$ after each such sweep. The probability that this stage will perform more than $K2^{k_F i}$ sweeps (and therefore will request more than $iK2^{k_F i}$ symbols from the prover) is  $\paren*{1-2^{-h_V i}}^{K2^{k_F i}}$. 
So the probability $p_1$ that $P^{C}$ will be accepted by employing \Cref{strat:1} is at most $1-p\paren*{1-2^{-h_V i}}^{K2^{k_F i}}$. 

\newcommand{\pagree}{\ensuremath{p_{\text{agree}}}}

If $P^{C}$ employs \Cref{strat:2}, it will be accepted if one of the  following events occur:
\begin{itemize}
    \item $V$ selects the CHECK-RULER stage, which accepts $P^{C}$. The probability of this event is at most $p$.
    \item $V$ selects the FREIVALDS stage, which reaches the same conclusion about the core as previously announced by $P^{C}$, after requesting  more than $iK2^{k_F i}$ symbols from the prover. Since $P^{C}$ can ``lie'' after some point during its transmission to direct $V$ to reach a conclusion consistent with its previous claim, an upper bound for the probability of this event is $(1-p)\paren*{\sfrac{1}{K}}$.
    \item $V$ selects the FREIVALDS stage, which reaches the same conclusion about the core as previously announced by $P^{C}$, after requesting no more than $iK2^{k_F i}$ symbols from the prover.  The probability of this event is not more than $(1-p)\pagree$, where we define $\pagree$ as the probability that those two conclusions agree. 
\end{itemize}

Assuming that $P^{C}$ is unable to separate \padlang{\squarelang} and \padlang{\subsquare}, there will be a set $W'_{P^C} \subseteq W$ containing infinitely many input strings for which it will send the correct answer with probability at most $\frac{1}{2}$ during the CLOCK stage of $V$. For any such input, $\pagree$, which is the sum of two terms corresponding to the events where $P^{C}$ and $V$ agree on the correct and wrong answer, respectively, is at most
\begin{equation*}
    \frac{1}{2}(1-\err_F)+\frac{1}{2}(\err_F)=\frac{1}{2},
\end{equation*}
where we recall that $\err_F$ is the (very small) probability that $V$ comes to the wrong conclusion about the core as a result of running Freivalds' algorithm. Therefore, the probability $p_2$ that  $P^C$ will be accepted  when it employs \Cref{strat:2} is at most $p+(1-p)\paren*{\frac{1}{K}}+\frac{1-p}{2}$ for infinitely many inputs. 

As one of the two ``pure'' strategies that we considered must yield the highest acceptance probability possible for $P^{C}$,\footnote{No probabilistic mixture of the two strategies can yield a higher acceptance probability.} this probability is bounded by 
\begin{equation*}
    \max\Set{1-p\paren*{1-2^{-h_V i}}^{K2^{k_F i}}, p+(1-p)\paren*{\frac{1}{K}}+\frac{1-p}{2}}.
\end{equation*}
One wishes  this probability to be as low as possible, so consider  setting the value of $K$ to a very large number, forcing $p_2$ to be below a value that is as near to $\frac{1+p}{2}$ as one desires.  For any given value of $K$, one can then set $h_V$ to a correspondingly large value that will force $p_1$ to be as near to $1-p$ as desired for all sufficiently long strings that form an infinite subset $W_{P^C}$ of $W'_{P^C}$.\footnote{$k_F$ has already been fixed by our choice of $c_F$ and $d_F$. We make use of the fact that $\lim_{x \to \infty} \paren*{1-2^{-Ax}}^{2^{Bx}}=1$ when $A>B>0$.} 
We fix  $p$ to the value $\frac{1}{3}$ that minimizes $\max\Set{1-p, \frac{1+p}{2}}$, and conclude that the probability that $P^{C}$ can be accepted by $V$ can be pulled down to any value above $\frac{2}{3}$, obtaining a success gap
\begin{equation*}
    \inf \Set{ \pr_{\acc}\paren*{V,P^{Q},w} | w \in W_{P^C} } - \sup \Set{ 1-\pr_{\rej}\paren*{V,P^C,w} | w \in W_{P^C} }
\end{equation*}
that can be set to be greater than any desired value $\Delta<\frac{1}{3}$. 

Note that $V$ runs in expected polynomial time only on input strings satisfying the promise that the core's length is logarithmic in the length of the input,\footnote{This is in contrast to the verifiers of \Cref{thm:paddedprotocol}, which satisfy the runtime requirement stated in \Cref{def:PoQ} for all possible input strings.} and is guaranteed to halt with probability 1, unlike the verifiers of \Cref{thm:paddedprotocol}.

$V$ can be said to  distinguish quantum provers from classical ones  only on the condition that no \tpfa{} can separate \padlang{\squarelang} and \padlang{\subsquare}. If that assumption is not true, a classical prover can achieve the same performance as the \tqfa{} in \Cref{fig:sqpro}. Of course, $V$ can also be seen as a framework enabling a party who has invented \emph{any} fast (classical or quantum) constant\-/space algorithm for solving the separation problem $\paren*{\padlang{\squarelang}, \padlang{\subsquare}}$ to demonstrate this to another party without revealing the source code.

\section{Concluding remarks}\label{sec:conc}

As we noted in \Cref{subs:pfaqfa}, there exist recognition problems which are known to be solvable in polynomial expected time by \tqcfa{}'s and at least exponential expected time by \tpfa{}'s. 
Presently, all known small\-/space \qtm{}'s that recognize such nonregular languages are  based on the same algorithmic technique~\cite{AW02,R20}, and the associated languages are essentially variants of $\mixedeq{}$, as described in that section.\footnote{All other small\-/space quantum algorithms mentioned in \Cref{subs:pfaqfa} have exponential runtime. In fact, Remscrim has shown~\cite{R21} that no sublogarithmic\-/space \qtm{} can recognize \pal{} in polynomial time.} This proven power difference between classical and quantum models  suggests a variant of the protocol described in \Cref{thm:paddedprotocol}, where the underlying problem is simply the recognition of such a language. However, neither Dwork and Stockmeyer's result~\cite{DS90}  establishing that any small\-/space \ptm{} recognizing a nonregular language must run in superpolynomial expected time, nor our generalization (in the \nameref{sec:generalizedDS}) of that theorem to machines which do not necessarily halt with probability $1$ is  strong enough
to guarantee that the resulting verifier will reject every classical prover with higher probability than the genuine quantum prover for that problem. Those theorems do not preclude a classical machine (say, a \tpfa{}) which recognizes $\mixedeq$ and runs within a polynomial time bound with probability $\sfrac{n-1}{n}$, and spends exponential time with probability $\sfrac{1}{n}$ for inputs of length $n$. If such a machine $P^C$ exists, it will be able to perform the recognition task with very high probability within some polynomially large time given by the verifier, and the argument used to demonstrate the success gap in the proof of \Cref{thm:paddedprotocol} will not work for a verifier faced with $P^C$. This line of thought about the search for more proof\-/of\-/quantumness protocols raises the following questions:

\begin{enumerate}[leftmargin=*]
    \item
        Is there a \tpfa{} that recognizes a nonregular language and runs within a polynomial time bound with probability $1-\oh{1}$ as a function of the input length?
    \item
        Do there exist interactive proof systems (like the ones presented in \Cref{thm:2dfass}) for languages other than \pal{} and \twin{} in \Cref{tab:languagesrecognizedinexp}?
    \item
        Which, if any, languages in \Cref{tab:languagesrecognizedinexp} (in addition to \pal{}, \twin{}, \mult{}, and ``word problems'', which are closely connected to \pal{}) yield  \dsfk{} problems when one considers separating their padded versions from those of their complements?
    \item
        Is $\TDFA[2] = \TDFA[2_s]$?
\end{enumerate}

In \Cref{sec:intro}, we mentioned the recent result~\cite{GRZ24} that every language in \BQL{} has an IPS through which a quantum logspace prover can prove the membership of a string in that language to a classical logspace verifier.\footnote{In fact, the protocol described in~\cite{GRZ24} is one\-/way, that is, the prover simply sends the verifier a single proof string, with no further interaction.} Although our work can be seen as a step towards a similar framework for the small\-/space case, the construction in~\cite{GRZ24} requires space to be at least logarithmic in time, and we conjecture that the answer to the following question is negative:

\begin{enumerate}[resume, leftmargin=*]
    \item
        Does there exist a general small\-/space IPS framework (\eg, for all languages in \BQTISP{t(n), s(n)} where $t(n)$ is some polynomial in $n$ and $s(n)\in \oh{\log \log n}$) through which a quantum $s(n)$\=/space prover can prove the membership of a string in a language that it can recognize to a classical  verifier with the same space bound? 
\end{enumerate}

\section*{Acknowledgments}

We thank Abuzer Yakary{\i}lmaz and  Wei Zhan for their helpful answers to our questions. 

\bibliographystyle{abbrvnat}

\bibliography{references} 




\clearpage



\begin{appendices}
\sectionfont{\LARGE}
\section*{Appendix}
\label{sec:generalizedDS}

\subsection*{A generalization of Dwork and Stockmeyer's theorem}


\begin{theorem}\label{thm:HALTWITH1ORLESS}
Let $M$ be a \ptm{} which recognizes a nonregular language such that for any input $w$ of length $n$, $M$ runs within space $\oh{\log \log n}$, halts with probability $h_w \leq 1$ within expected time $T(n)$, and fails to halt with the remaining probability $1-h_w$. Then, for every $b < 1$,
\begin{equation*}
    \log \log T(n) \geq (\log n)^b
\end{equation*}
for infinitely many $n$. In particular, $T(n)$ is not bounded above by any polynomial in $n$.
\end{theorem}
\begin{proof}
We start with a quick recap of Dwork and Stockmeyer's proof of the case (\cite[Theorem~5.1]{DS90}) where the \ptm{} halts with probability 1:

For any language $\langL$ and number $n>0$, two strings $w$ and $w'$ are said to be  \emph{$n$\-/dissimilar}, written $w \not\sim_{\langL,n} w'$, if $\abs{w} \leq n $, $\abs{w'} \leq n$, and there exists a \emph{distinguishing string} $v$ with $\abs{wv} \leq n$, $\abs{w'v} \leq n$, and $wv \in \langL \iff w'v \notin \langL$. Let $N_{\langL}(n)$ be the maximum $k$ such that there exist $k$ distinct strings that are pairwise $\not\sim_{\langL,n}$. It is known~\cite{SB96}  that, if $\langL$ is nonregular, then   $N_{\langL}(n) \geq \frac{n+3}{2}$ for infinitely many $n$.

It is convenient to adopt the following new conventions regarding $M$:
$M$ starts in a special state $s_1$ (which will only be used for the initial right-to-left pass) with the input head on the right endmarker $\rend$, moves the input head to the left endmarker without moving the work tape head, and switches to a different state when it arrives at $\lend$. $M$ keeps track of the rightmost non\-/blank symbol on the work tape throughout its execution, and performs a subroutine that ``cleans'' the work tape by replacing all non\-/blank symbols by blanks and positioning the work head on the leftmost end of the  tape immediately before it accepts or rejects.

Let us call an instantaneous description of the collection of the internal state, work tape content, and work head position of $M$ a \emph{megastate} of $M$. (The familiar notion of a \emph{configuration} of $M$ then corresponds to a pair consisting of a megastate of $M$ and the position of its input tape head.)
Since $M$ has space complexity $\oh{\log \log n}$, $c(n)$, the number of distinct megastates that it can attain while running on an input of length $n$, grows more slowly than $(\log n)^a$ for any constant $a>0$. We impose some ordering from 1 to $c(n)$ on the megastates of $M$, with megastate 1 corresponding to $s_1$ and a ``clean'' work tape containing all blank symbols with the work head positioned on the left end. Megastates $c(n)-1$ and $c(n)$ correspond to the  reject and accept states, respectively, both with similarly clean work tapes.

As Dwork and Stockmeyer describe~\cite{DS90}, there exists a Markov chain $P_{M,xy}$ with $2c(n)$ states that models the computation of $M$ on the concatenated string $xy$ for any two strings $x,y\in\Sigma^*$, where $\abs{xy}=n$. 
For $1 \leq j \leq c(n)-1$, state $j$ of $P_{M,xy}$ corresponds to $M$ being in the configuration
with megastate $j$ and the input head on the last symbol of $\lend x$, and state $c(n) +j-1$
 corresponds to $M$ being in the configuration with megastate $j$ and the input head on
 the first symbol of $y\rend$. (State  1 of $P_{M,xy}$ thus corresponds to a configuration that $M$ will be in with probability 1 shortly after the beginning of its computation.) 
Importantly, state $2c(n)-1$ of $P_{M,xy}$ corresponds to a disjunction of rejection, infinite loop
 with the input head never leaving the region $\lend x$, and infinite loop within the region $y \rend$. State $2c(n)$ of $P_{M,xy}$
 corresponds to acceptance.  This Markov chain is ``faithful'' to $M$, in the sense that each transition probability of $P_{M,xy}$ equals the probability of the (multi\-/step) transition between the corresponding configurations of $M$ when running on $xy$.   The probability that  $P_{M,xy}$ is absorbed in state $2c(n)$ when started
 in state 1 equals the probability that $M$ accepts $xy$. Since $M$  halts with certainty in the case in~\cite{DS90} (where $h_w=1$ for all $w$), the chain is absorbed in state $2c(n) - 1$ with the remaining probability. The relationship between the computation of $M$ and $P_{M,xy}$  guarantees that the Markov chain's expected time to absorption (into one of those two states) is at most $T(n)$ as well.

One then assumes that $T(n)$ is smaller (\ie, $M$ runs faster) than dictated by the lower bound in the theorem statement. This assumption and the aforementioned lower and upper bounds for $N_{\langL}(n)$ and $c(n)$, respectively, lead to the conclusion that, for the language $\langL$ recognized by $M$ and  for sufficiently large $n$, there exist two pairwise $ \not\sim_{\langL,n} $ strings $w$ and $w'$, with distinguishing string $v$, such that the Markov chains $P_{M,wv}$ are $P_{M,w'v}$ are so close (according to a notion of closeness defined in~\cite{DS90})  to each other that $M$ must accept either both or none of the strings $wv$ and $wv'$, contradicting their $n$\=/dissimilarity, and concluding the proof. The assumption that $T(n)$ (\ie, the absorption times for both Markov chains) is ``small'' plays a critical role in this final step.  

We now show how to modify the proof above to also work for the case where $M$ may run forever on any input string $w$ with some nonzero probability $1-h_w$. Call a non\-/halting configuration $C$ ``looping'' if  $M$'s probability of halting eventually is 0 when it is started from $C$. Note that $M$ attains at least one looping configuration with nonzero probability while running on input $w$ if and only if 
$M$ halts with probability less than 1 on $w$. Since $M$ recognizes a language, there is a probability greater than $\frac{1}{2}$ that its computation will consist of a finite sequence of non\-/looping configurations followed by a halting configuration. By the theorem 
statement, the expected length of such a sequence is at most $T(n)$. A looping computation  can occur with  probability less than $\frac{1}{2}$, and consists of a finite sequence of non\-/looping configurations, followed by an infinite sequence of looping configurations.

If $M$ can loop with nonzero probability on an input string $xy$, the corresponding Markov chain  $P_{M,xy}$ constructed using the procedure described above is no longer guaranteed to have a finite, let alone ``small'',  expected absorption time. The given construction does map  infinite computations of $M$ where the input head does not leave one of the $\lend x$ and $y \rend$ regions to finite paths that end in state $2c(n)-1$ in the chain. However, if $M$ shuttles its input head back and forth forever between the two regions, the corresponding Markov chain also ``runs'' forever.

We solve this problem by adding a ``postprocessing'' step in the construction of $P_{M,xy}$. Let $p_{i,j}$ denote the probability of the transition from state $i$ to state $j$ set by the previously mentioned procedure. We simply ``rewire'' the chain by diverting every transition to a state corresponding to a looping configuration to the trap state $2c(n)-1$:

\begin{turingenum}
    \titem{For each state $j$ that corresponds to a looping configuration of $M$:}
    \ttitem{For each $i$:}
    \tttitem{$p_{i,2c(n)-1}\leftarrow p_{i,2c(n)-1}+p_{i,j}$}
    \tttitem{$p_{i,j}\leftarrow 0$}
\end{turingenum}

The modified chain  still models the acceptance probability of $M$ faithfully. All non\-/halting computations of $M$ are modeled by finite paths in the chain. Note that these paths correspond to sequences of non\-/looping configurations of $M$, which also appear in halting computations, so their lengths are accounted for in the expected runtime $T(n)$ of the ``halting part'' of the machine. We conclude that the modified chain's expected absorption time is also bounded by $T(n)$. The rest of the proof for the general case is identical to that of the case for $h_w=1$.
\end{proof}

\end{appendices}

\end{document}